\documentclass[usenatbib,useAMS]{mn2e}
\usepackage{epsfig}
\usepackage{times}
\usepackage{amsmath}

\usepackage{verbatim}

\newcommand{\simgt}%
        {\,\hbox{\lower0.6ex\hbox{$\sim$}\llap{\raise0.6ex\hbox{$>$}}}\,}
\newcommand{\simlt}%
        {\,\hbox{\lower0.6ex\hbox{$\sim$}\llap{\raise0.6ex\hbox{$<$}}}\,}

\title[]{Multi-dimensional modelling of X-ray spectra for 
AGN accretion-disk outflows III: application to a hydrodynamical simulation}
\author[Sim et al.]{S. A. Sim$^1$, D. Proga$^{2}$
L. Miller$^3$, K. S. Long$^4$, T. J. Turner$^{5}$\\
$^{1}$Max-Planck-Institut f\"{u}r Astrophysik,
Karl-Schwarzschildstr. 1, 85748 Garching, Germany\\
$^{2}$Department of Physics, University of Nevada, Las Vegas, NV 89154\\
$^{3}$Dept. of Physics, University of Oxford, Denys Wilkinson
Buiding, Keble Road, Oxford OX1 3RH, U.K.\\
$^{4}$Space Telescope Science Institute, 3700 San Martin Drive,
Baltimore, MD 21218, U.S.A\\
$^{5}$Dept. of Physics, University of Maryland Baltimore County, 1000 Hilltop Circle, Baltimore, MD 21250, U.S.A\\
}

\date{\today}

\pagerange{\pageref{firstpage}--\pageref{lastpage}}
\pubyear{}
\begin{document}
\maketitle
\label{firstpage}

\begin{abstract}
We perform multi-dimensional radiative transfer simulations to compute
spectra for a hydrodynamical simulation of a 
line-driven accretion disk wind from an active galactic nucleus.  The 
synthetic spectra confirm expectations from parameterized models that a 
disk wind can imprint a wide variety
of spectroscopic signatures including narrow absorption lines, broad
emission lines and a Compton hump. 
The formation of these features is complex with
contributions originating from many of the different structures
present in the hydrodynamical simulation. 
In particular, spectral features are shaped both by gas in 
a successfully launched outflow and in
complex flows where material is lifted out of the disk plane but ultimately
falls back.
We also confirm that the strong
Fe~K$\alpha$ line can develop a weak, red-skewed line wing as a result
of Compton scattering in the outflow. 
In addition, we demonstrate that X-ray
radiation scattered and reprocessed in the flow has a pivotal part in
both the spectrum formation and determining the ionization conditions
in the wind. We find that scattered radiation is rather effective in
ionizing gas which is shielded from direct irradiation from the
central source. This effect likely makes the successful launching of a
massive disk wind somewhat more challenging and should be considered
in future wind simulations.
\end{abstract}

\begin{keywords}
radiative transfer --  methods: numerical -- galaxies: active --
X-rays: galaxies
\end{keywords}

\section{Introduction}
\label{sect_intro}

Estimates for the black hole mass ($M_{\mbox{\scriptsize bh}}$) and bolometric luminosity
($L_{\mbox{\scriptsize bol}}$) of the
most luminous active galactic nuclei (AGN) suggest that they
are accreting close to the Eddington rate,
$\dot{M}_{\mbox{\scriptsize Edd}}$. 
Under such
conditions, radiation forces should play a role in
determining the properties of the accretion disk atmosphere. In
particular, since much of the luminosity of the inner regions of the
disk is radiated in the ultraviolet (uv) part of the spectrum, a
significant radiative force should arise owing to the large number of
line transitions which occur in this spectral region. By analogy to hot
star atmospheres, it is to be expected that radiation pressure due
to ultraviolet line driving can exceed that due to electron scattering
by factors up to $M_{\mbox{\scriptsize max}}\sim 10^3$ \citep{castor75,stevens90,proga00} under optimal conditions. Thus the radiation force due to spectral
lines might be expected to affect the structure of a uv-bright accretion
disk for luminosities as small as $\sim M_{\mbox{\scriptsize max}}^{-1}$
$L_{\mbox{\scriptsize Edd}}$. 

In luminous hot stars
line-driving by the uv radiation field is able to launch
fast winds with terminal velocities comparable to the escape speed from
the stellar photosphere
(e.g. \citealt{castor75,friend86,pauldrach86}). 
For AGN accretion
disks, launching of a wind
is hindered by the large luminosity of ionizing X-ray radiation
produced in the vicinity of the central black hole. Line-driving
becomes rapidly less efficient as the ionization state of the gas is
raised (see e.g. \citealt{stevens90,arav94}). As a consequence, for luminosities
around $0.001
\simlt L/L_{\mbox{\scriptsize Edd}} \simlt 0.1$, line-driving is unable to launch a
successful wind but should give rise to a topologically
complex ``failed wind'' in which material is lifted out of the disk
plane but, due to over-ionization by the X-ray radiation, does not reach escape speed (\citealt{proga04}; hereafter PK04,
see also \citealt{proga00,proga05}).
Above $L \sim 0.1~L_{\mbox{\scriptsize Edd}}$,
however, numerical simulations suggest that 
even the ionizing X-ray flux may be unable to fully suppress the
launching of a line-driven disk wind. 
In particular, PK04 
performed radiation-hydrodynamical
simulations for an accretion disk around a supermassive black hole
($M_{\mbox{\scriptsize bh}} = 10^8$~M$_{\odot}$).
They investigated
the case of $L \sim 0.5 L_{\mbox{\scriptsize Edd}}$ assuming that
10 per cent of the luminosity was radiated as X-rays from a centrally
concentrated source. This central source could ionize the gas but 
was not included as a source of radiation pressure -- 
the line driving force was supplied by the uv radiation of the
accretion disk.
Their simulations
predicted that, despite the ionizing
effects of the X-ray radiation, a significant mass flux was
successfully accelerated that formed an outflow with terminal velocity
comparable to the escape speed from the inner accretion disk. 

The flow
structure found by PK04 
is complex and significantly time-variable but it is persistent and
characterized by three regimes: a cold dense equatorial outflow, a hot
diffuse polar flow and a thin layer of moderate density rapidly
outflowing warm material at intermediate inclination. The typical mass-loss
rate of the flow was $\sim 0.1 - 0.2$~M$_{\odot}$~yr$^{-1}$, this being
dominated by the warm, rapid outflow component. In addition, the
simulation predicts a dense ``failed wind'' region above the inner
disk which 
plays a pivotal part in
determining the wind structure owing to both the ram pressure it
exerts and its potential role in shielding the gas in the outer
regions of the flow from the ionizing X-ray radiation. We note, 
however, that PK04 assumed that all the X-ray radiation is
emitted by a point source and they neglected scattering or
reprocessing of X-rays in the flow. As will be discussed later, this
likely overestimates the ease with which the ``failed wind'' can
shield the outer parts of the flow.

Given the
geometrical and kinematic complexity of the PK04 wind structure,
predicting its spectroscopic signatures is challenging. But given the
large mass-loss rate of the warm stream and the significant covering fraction of the ``failed wind'', it
is certainly expected that observational signatures will arise. 
Since a large fraction of the material around the inner parts of the
flow should be strongly ionized, the clearest signatures of
the absorption and emission from the wind structure may be expected to
manifest in the X-ray band.
\cite{schurch09}
performed radiation transfer calculations for snapshots of the wind
structure obtained by PK04, again assuming that all X-rays are emitted
by a single point source. They confirmed that, depending on
the observer's orientation, the wind structures could
strongly absorb X-ray radiation and therefore imprint observable
signatures. Although an important step, the
\cite{schurch09} study suffers from the shortcoming that the radiative
transfer code used ({\sc xscort}; \citealt{schurch07}) is only
  one-dimensional. Thus, although they could study the absorption
  properties of the disk wind, they did not realistically address the
  role of the wind in producing/shaping emission features.

We have developed a multi-dimensional Monte Carlo radiative transfer
code which is able to compute orientation-dependent spectra for
wind models taking proper account of scattering and reprocessing of
radiation in the outflow (\citealt{sim05b}, \citealt{sim08}; hereafter
Paper~I, \citealt{sim10}; hereafter Paper~II). Applying this code
to simply-parametrized disk wind geometries has shown that
scattered/reprocessed radiation has a very important role in the
formation of the X-ray spectrum and that outflow models may have the
capacity to explain a wide variety of X-ray spectroscopic features
(see Papers I and II). In particular, highly blue-shifted absorption
features, as have been reported in a number of luminous AGN (see
section 6.1 of \citealt{turner09} for a recent review), appear
for a subset of inclination angles while Fe~K$\alpha$
emission is predicted for all observer orientations. The line emission
is typically broad and can develop somewhat red-skewed line wings
(Papers I and II)
owing to the effects of Compton scattering in the flow
\citep[see e.g.][]{laming04,laurent07,titarchuk09}.

To date, however, we have only applied our Monte Carlo code to
parametrized outflow models with imposed mass-loss rates and flow
geometries. 
Although such simple models provide a
convenient means to investigate the formation of the X-ray
spectrum in an outflow, they do not adequately represent the
full complexity of the wind structures found in the
radiation-hydrodynamical simulations mentioned above.
In particular, they adopt a very simple velocity law (which is
not dynamically consistent with any particular
launching/acceleration mechanism)
and assume that
all the material that is lifted out of the disk plane is part of a
flow that successfully escapes from the system.
Thus, at best, the parametrized wind models might represent
components similar to the warm outflow in the PK04 simulation but they
neglect any other structures which accompany it (e.g. the ``failed wind'').
It is therefore important to
investigate whether there are characteristic differences
between the X-ray spectra obtained from simplified wind models and
dynamically self-consistent flow models since these are relevant to
the quantitative interpretation of observations.

Here we take the step of connecting our studies of the formation of
X-ray spectra in multi-dimensional flows (Papers I and II) 
to a physically motivated
disk wind model (PK04).
The objectives of this study are two fold. First, by computing synthetic X-ray
spectra for a hydrodynamical simulation
we can more reliably investigate which spectral
features can plausibly form in disk winds and assess
how well synthetic spectra from parametrized models (such as those
used in Paper II) correspond to what might be expected from
more realistic flow structures. Second, by using a radiative transfer
scheme that readily incorporated true absorption, scattering and
reprocessing of radiation in the wind, we can begin to assess the
importance of scattered radiation for the physical processes
at work in a line-driven wind. 
We begin, in Section~\ref{sect:method} with a
description of the setup of our simulations. The results are presented
in Section~\ref{sect:results} and their implications discussed in
Section~\ref{sect:discussion}.

\section{Model setup}
\label{sect:method}

\subsection{Radiation-hydrodynamical wind simulation}

To provide a set of self-consistent gas properties for our Monte Carlo
radiative transfer simulations,
we use the hydrodynamic accretion disk wind simulations presented by PK04.
The hydrodynamical simulations assume a $10^8$~M$_{\odot}$ black hole accreting
at 1.8~M$_{\odot}$~yr$^{-1}$. These parameters correspond to an
accretion rate of $0.5~\dot{M}_{\mbox{\scriptsize Edd}}$ assuming a
radiative efficiency of 6~per cent.
The gas is assumed to
accrete through an optically thick, geometrically thin standard disk.
The disk thermal uv radiation drives the wind. The simulations also include
Comptonized continuum radiation from a central source which represents
the emission from the inner parts of the accretion disk that are not covered by
the computational domain.
This component is not included in the line-driving calculation for the wind,
because in the X-ray band there are relatively few
line transitions.
However, the central X-ray component
has a major effect on the derived photoionization state.
The properties of the outflow are calculated over the radial range 
60 -- 3000 $r_g$ (where $r_{g} = G M_{\mbox{\scriptsize bh}} / c^{2}$ is the
gravitational radius of the black hole).

To solve the equations of gas dynamics, PK04 used the {\sc zeus} code
which has been tested extensively (Stone \& Norman 1992).
Their version of the code has been modified to implement the radiation force,
and radiative heating and cooling.
PK04 approximated the radiative acceleration due to lines
using a modification of the method developed by
\cite{castor75}.
The radiative driving due to line scattering is treated using the generalized
Sobolev approximation. This assumes the transport in the line core
is dominated by the local velocity gradient.
The coupling between the
dynamics and the radiation field is provided by radiation pressure and by
thermal pressure due to radiative heating.  The radiation pressure depends
on the gas opacity, which in turn depends on the ionization balance.
The effects of X-ray ionization on the radiation force due to lines is
calculated using the procedure outlined by \cite{stevens90}.
In this procedure, the ``force multiplier'' of \cite{castor75}
(the ratio of the full radiation
force to that due solely to electron scattering) is suppressed by
a factor which depends on the ionization parameter and temperature.
The effects of heating by radiation are also taken into account
by these calculations.  Specifically,
PK04 calculated the gas temperature assuming that the gas is optically thin
to its own cooling radiation.  Thus the net cooling rate depends on
the density,
the temperature, the ionization parameter,
and the characteristic temperature of the X-ray radiation. In this case it
is possible to fit analytical formulae to the heating and cooling rate
obtained from detailed photoionization calculations for various gas
parameters.
PK04 used a fit to photoionization
calculations obtained by \cite{blondin94} who included Compton heating/cooling,
X-ray photoionization heating/recombination cooling,
bremsstrahlung and line cooling.

\begin{table*}
\caption{Parameters for the simulation.}
\label{tab:pars}
\begin{tabular}{ll}\\ \hline
Parameter & Value \\ \hline \hline 
mass of central object, $M_{\mbox{\scriptsize bh}}$ & $10^8$ M$_{\odot}$ \\
source luminosity, $L_{\mbox{\scriptsize bol}}$ & $6.5 \times 10^{45}$
ergs~s$^{-1}$ ($\sim 0.5
L_{\mbox{\scriptsize Edd}}$)\\
X-ray source luminosity (2 -- 10 keV), $L_{X}$ & $2.7 \times 10^{44}$
ergs~s$^{-1}$ ($\sim 0.02
L_{\mbox{\scriptsize Edd}}$) \\
source power-law photon index, $\Gamma$ & $2.1$ \\
range of source photon energies in simulation & 0.1 -- 511~keV \\
size of primary emission region, $r_{er}$ & $6 r_g = 8.8 \times 10^{13}$ cm \\
inner radius of disk,$r_{d}$ & $6 r_g = 8.8 \times 10^{13}$ cm \\
inner radius of simulated wind region, $r_{\mbox{\scriptsize min}}$ & 60 $r_g = 8.8
\times 10^{13}$ cm \\
outer radius of simulated wind region, $r_{\mbox{\scriptsize max}}$ & 3000
$r_g = 2.6 \times 10^{16}$ cm  \\
outer radius of Monte Carlo Rad. Trans. simulation grid & {$5 \times 10^{16}$ cm}\\
3D Cartesian Rad. Trans. grid cells & {$256 \times 256
  \times 256$}\\
2D radiation-hydrodynamics polar grid zones & {$100 \times 140$}\\
\hline
\end{tabular} \\
\end{table*}

\subsection{Monte Carlo radiative transfer simulations}

The hydrodynamical simulations described above provide the
time-dependent density and velocity structure of the disk wind. 
To compute synthetic spectra from these, we have performed radiative
transfer simulations using the Monte Carlo
code described in Papers I and II.

For the radiative transfer simulations, the Monte Carlo code was
modified to accept a
generalized axisymmetric wind with density and velocity specified via
input data. As for the parametrized wind models adopted in Papers I
and II, these input data define a 2D grid of wind properties (in this
case, with properties depending on polar coordinates $r$ and
$\theta$). The mass density in
each wind grid cell is assumed to be uniform and taken from the
hydrodynamical model. The three components of
velocity ($v_{r}$, $v_{\theta}$ and $v_{\phi}$) at every point in the
wind are obtained by linear interpolation (in $r$ and $\theta$) between the values which
are provided at the boundaries of each wind grid cell. This
interpolation is necessary since use of the Sobolev approximation for
line transitions 
requires that the velocity is everywhere a smooth function. 

In certain regions, particularly close to the disk plane where the
transition between the disk atmosphere and the outflow lies, the hydrodynamical
simulations predict rather high densities. In some cases, this makes
certain grid cells very optically thick. Physically, any X-ray
photons which penetrate deep within these optically thick layers will
be thermalized and contribute to heating of the disk atmosphere. In
the parametrized models considered in Paper~II, this sink of
X-ray photons was roughly accounted for by assuming that all Monte
Carlo quanta which reach the $xy$-plane in the simulation strike the
optically thick disk and are then lost. That approach neglects
reflection by the disk but avoids the need to track the Monte
Carlo quanta as they propagate through very optically thick
material. In the more realistic model considered here, however, there
is no sharply-defined boundary between the disk atmosphere and
the wind. Thus reflection by the dense material at the very base of
the outflow is automatically included in the simulations. However,
this introduces additional computational challenges. In particular,
quanta occasionally propagate deep into the optically thick regions
where they interact many times and have very low probability of 
reemerging without being thermalized (and therefore lost to the X-ray
regime since the expected temperature of the disk atmosphere is
relatively low). To deal with these quanta, a cut
is introduced whereby the flight paths of Monte Carlo quanta which propagate sufficiently
deep into the base of the wind are terminated and it is assumed that
these quanta contribute nothing further to the X-ray spectra or
ionization state of the wind. A very conservative cut has been applied
in the simulations presented here: quanta are terminated if they
penetrate deep enough into the base of the wind that the Compton 
optical depth to re-emerge from the wind is greater than twenty in all directions.

As described in Paper~II, the 
ionization and thermal structure of the wind are determined
iteratively from the radiation field properties and the assumptions of
ionization and thermal equilibrium. 
The ionization balance accounts for photoionization (including K-/L-shell
ionization followed by ejection of Auger electrons), collisional
ionization, radiative recombination and di-electronic recombination.
Heating by photoionization, Compton down-scattering of X-ray photons
and free-free absorption are balanced against bremsstrahlung, Compton
cooling by low energy disk photons, bound-free recombination,
bound-bound line transitions and cooling due to adiabatic expansion to
estimate the local temperature. Note that our treatment of Compton
cooling remains particularly approximate since we do not simulate the
transport of the accretion disk photons in detail (see Paper~II).
The assumptions of ionization/thermal equilibrium are not
perfectly valid in regions of the wind where the density is
sufficiently low that the 
recombination/cooling timescales are long compared to
the flow timescale. However, it alleviates the need to 
introduce explicit time-dependence in the calculation -- a major
computational saving. In general, we would expect that the most
important consequence of departures from equilibrium is that the
ionization state becomes frozen-in in the outer, low-density flow regions
since recombination may become too slow to maintain local ionization
equilibrium. As we shall show below, however, the
equilibrium assumption already leads to near fully-ionized
conditions in much of the outer wind -- thus accounting for any
further effective reduction to the recombination rate is unlikely to
dramatically alter the typical ionization state.

The radiative transfer simulations were performed using the same set
of atomic data as described in Paper~II.
This includes the K-shell ions of the astrophysically abundant
metals, the L-shell ions of the important second and third row
elements and the highest few M-shell ions of Fe and Ni.
Note that this data set does not include
M-shell ions of iron below Fe~{\sc x} -- thus when the
computed ionization state favours this ion it is likely that the true
degree of ionization is lower. 
Similarly,
a lower limit of
$\log T_{e}[\mbox{K}] = 4.0$ is imposed
in the radiative transfer simulations since (i) the atomic data used is
likely inadequate to describe the cooling to lower temperatures and
(ii) the uv radiation of the disk -- which is neglected here -- should 
prevent the inner regions of the outflow from dropping to
significantly lower temperatures.

\subsection{Simulation parameters}

\begin{figure*}
\epsfig{file=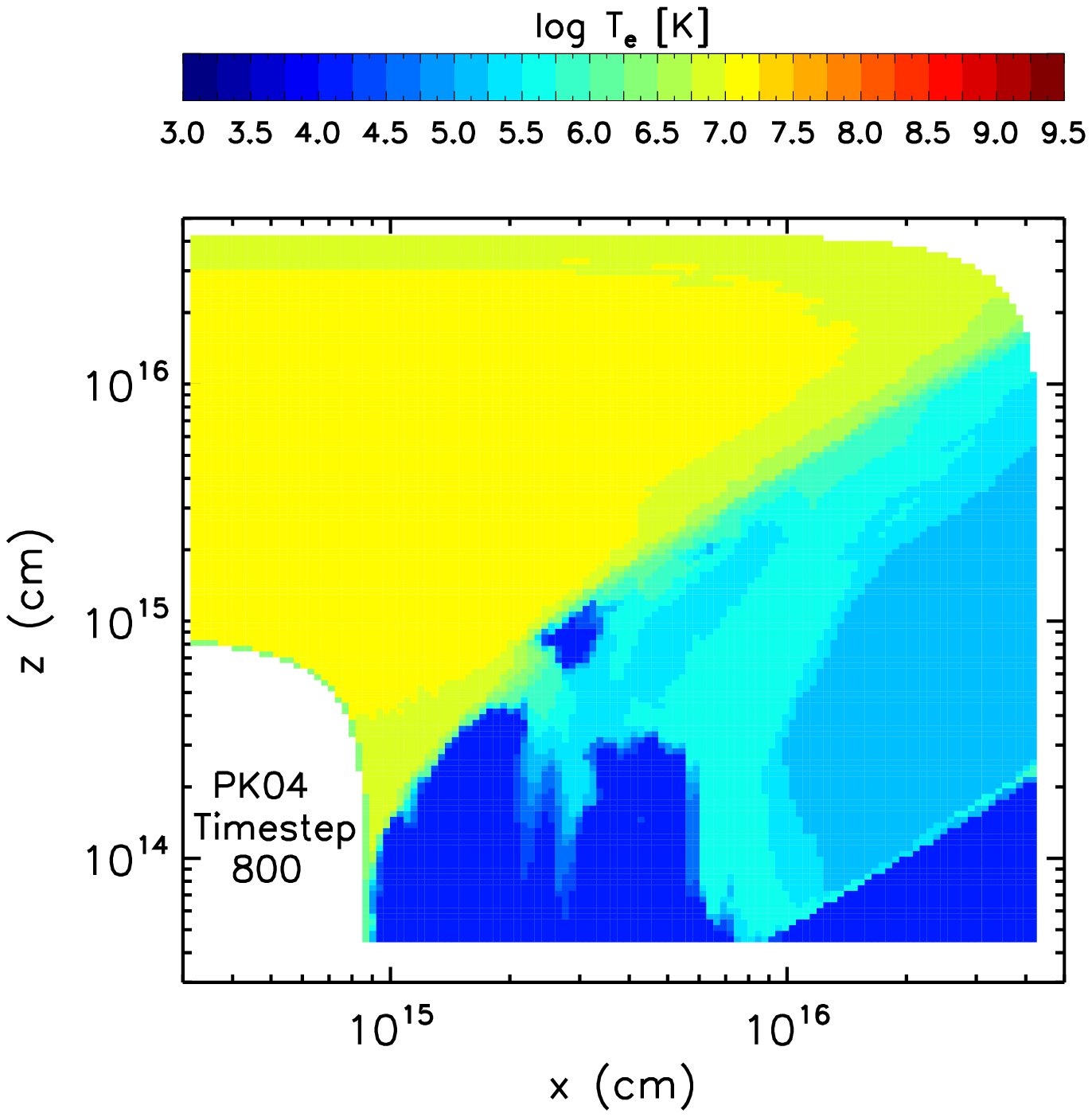, width=7cm,angle=90}
\epsfig{file=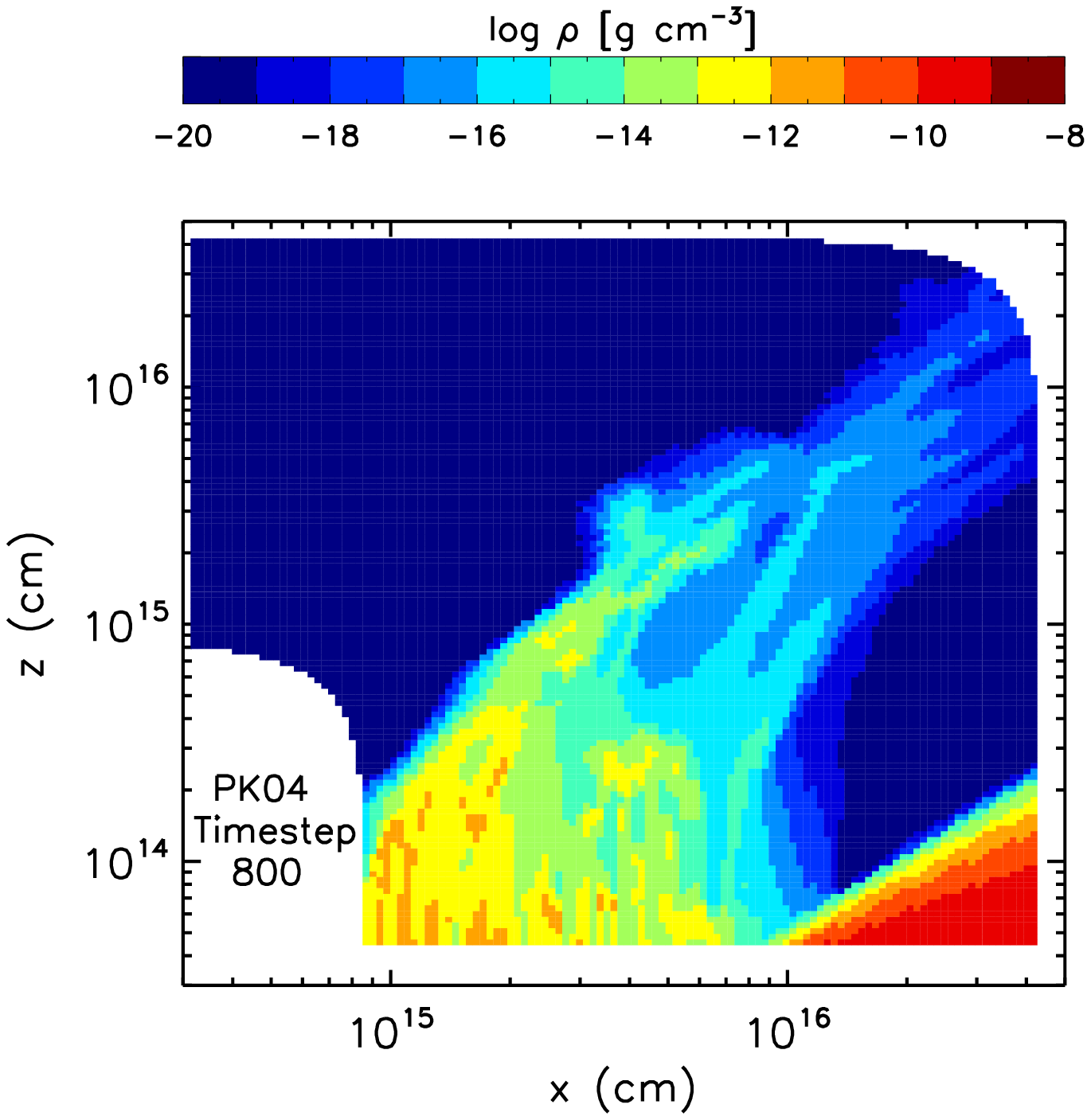, width=7cm,angle=90}\\
\epsfig{file=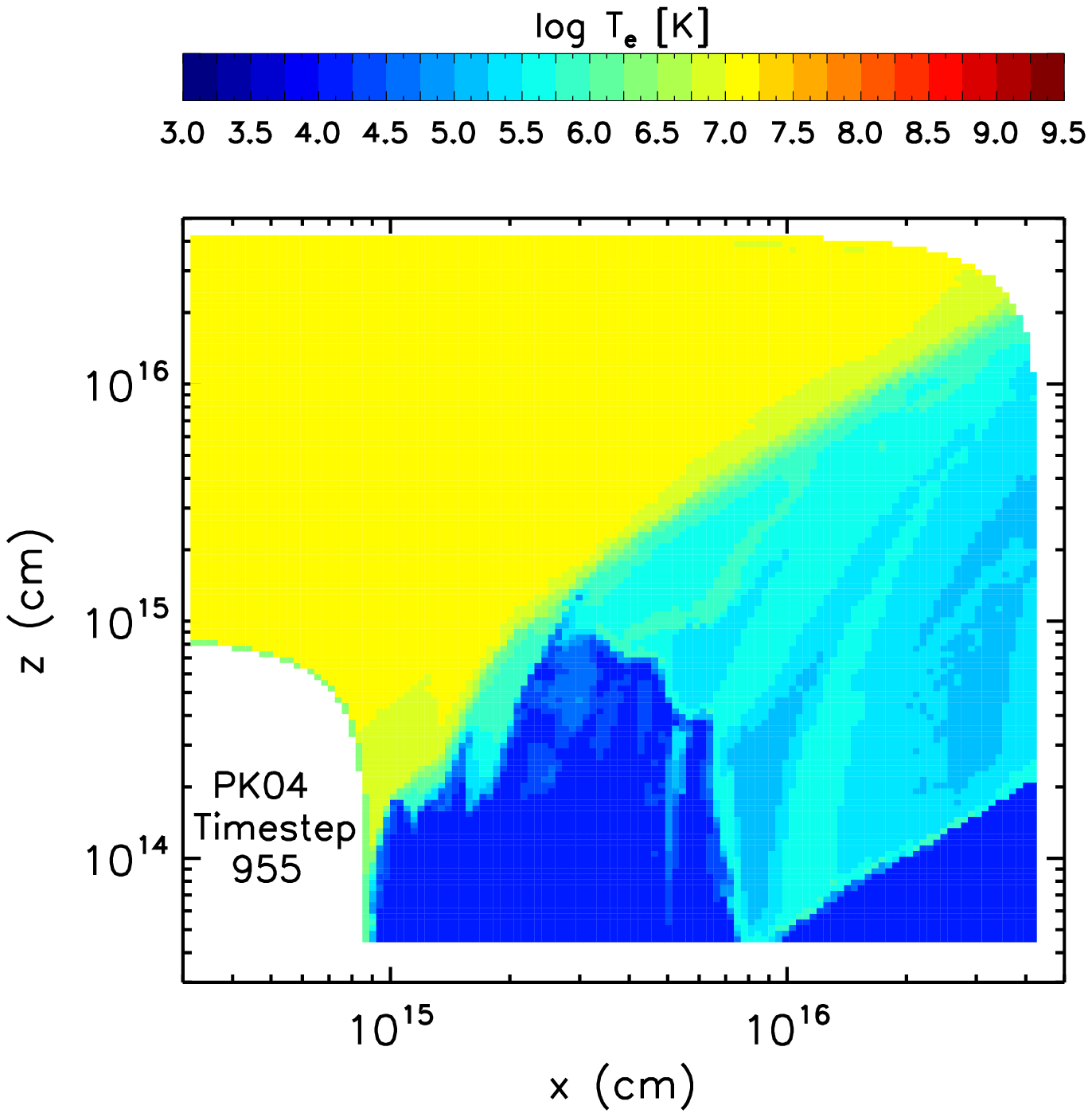, width=7cm,angle=90}
\epsfig{file=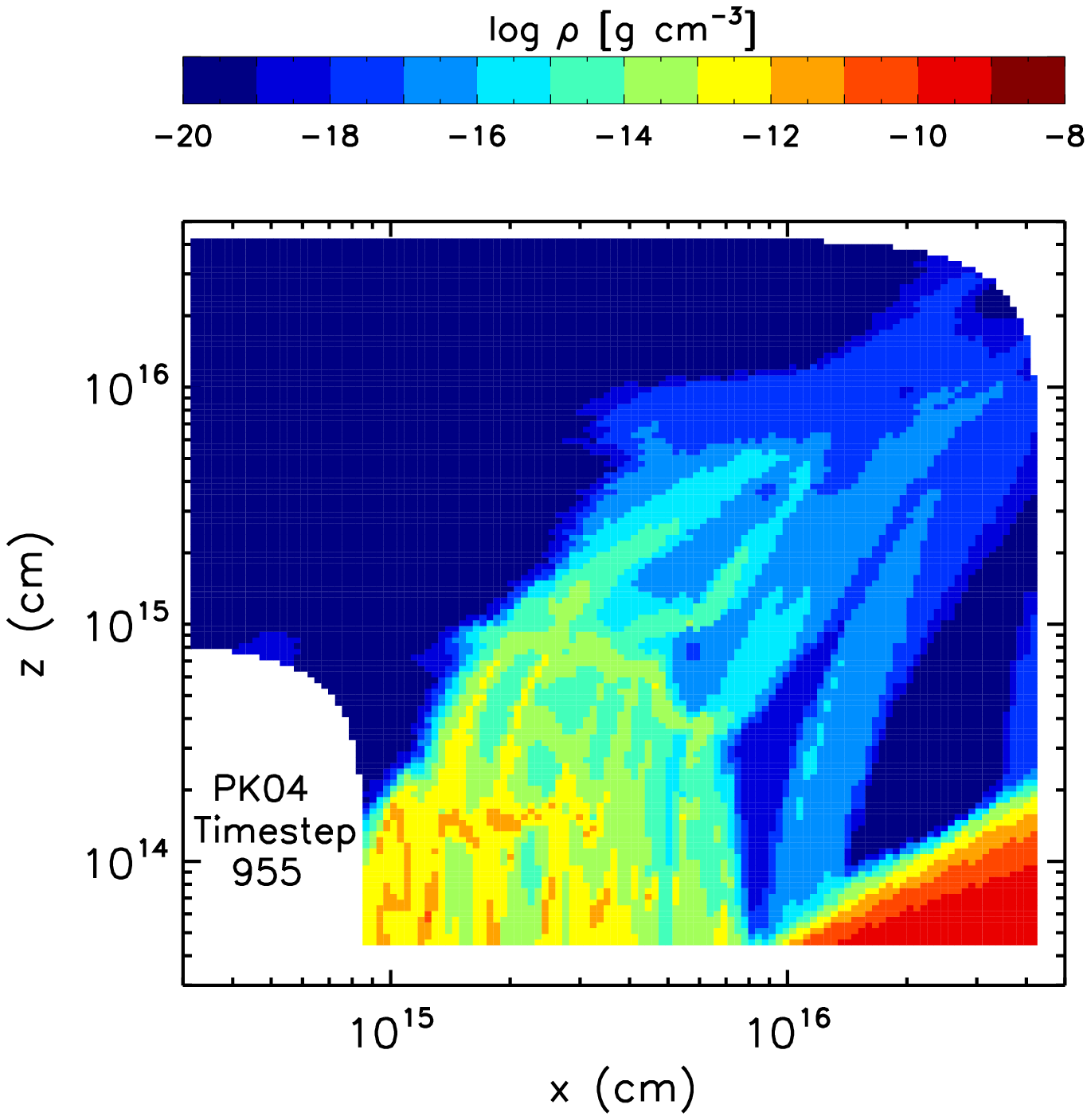, width=7cm,angle=90}\\
\epsfig{file=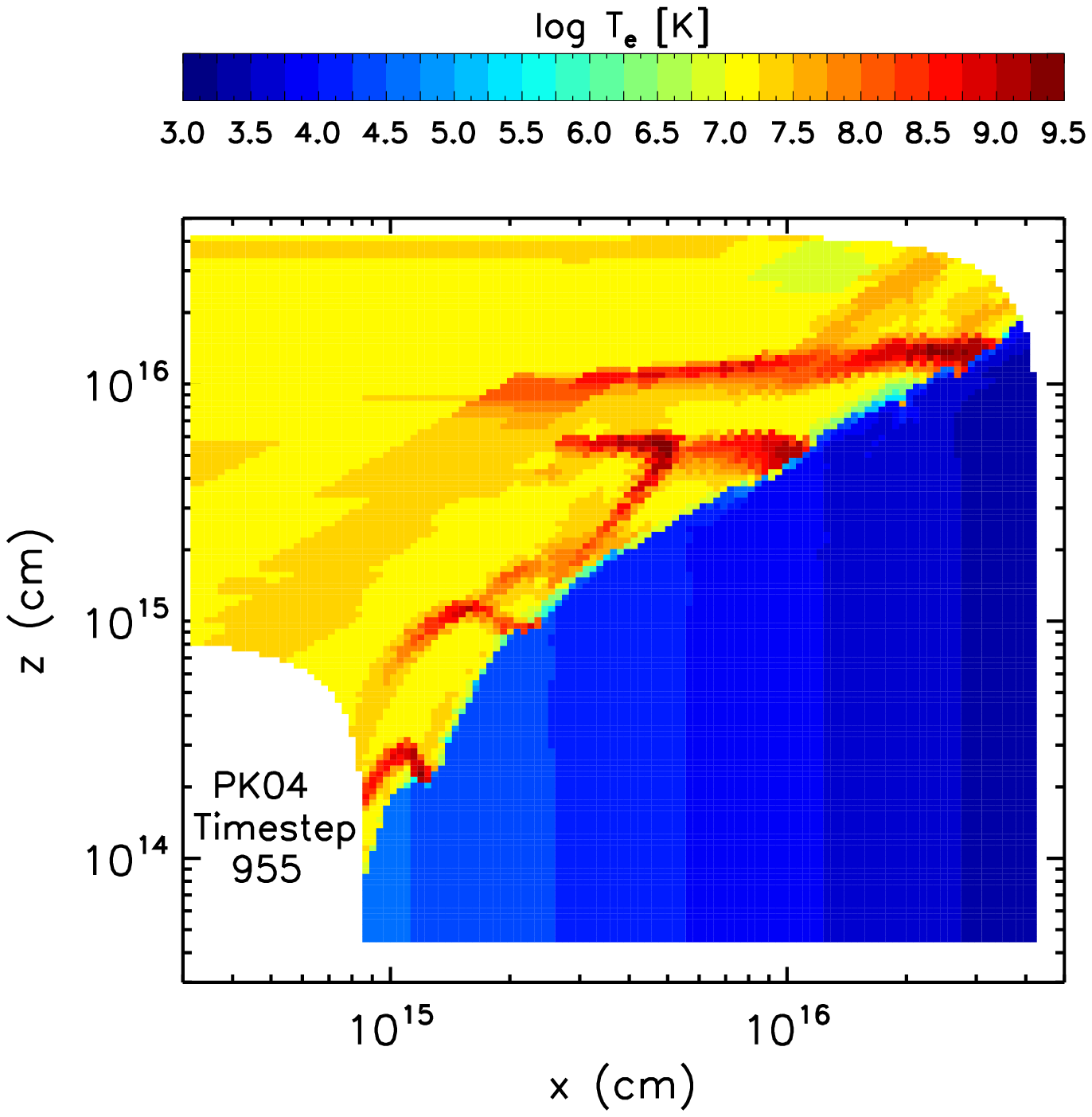, width=7cm,angle=90}
\epsfig{file=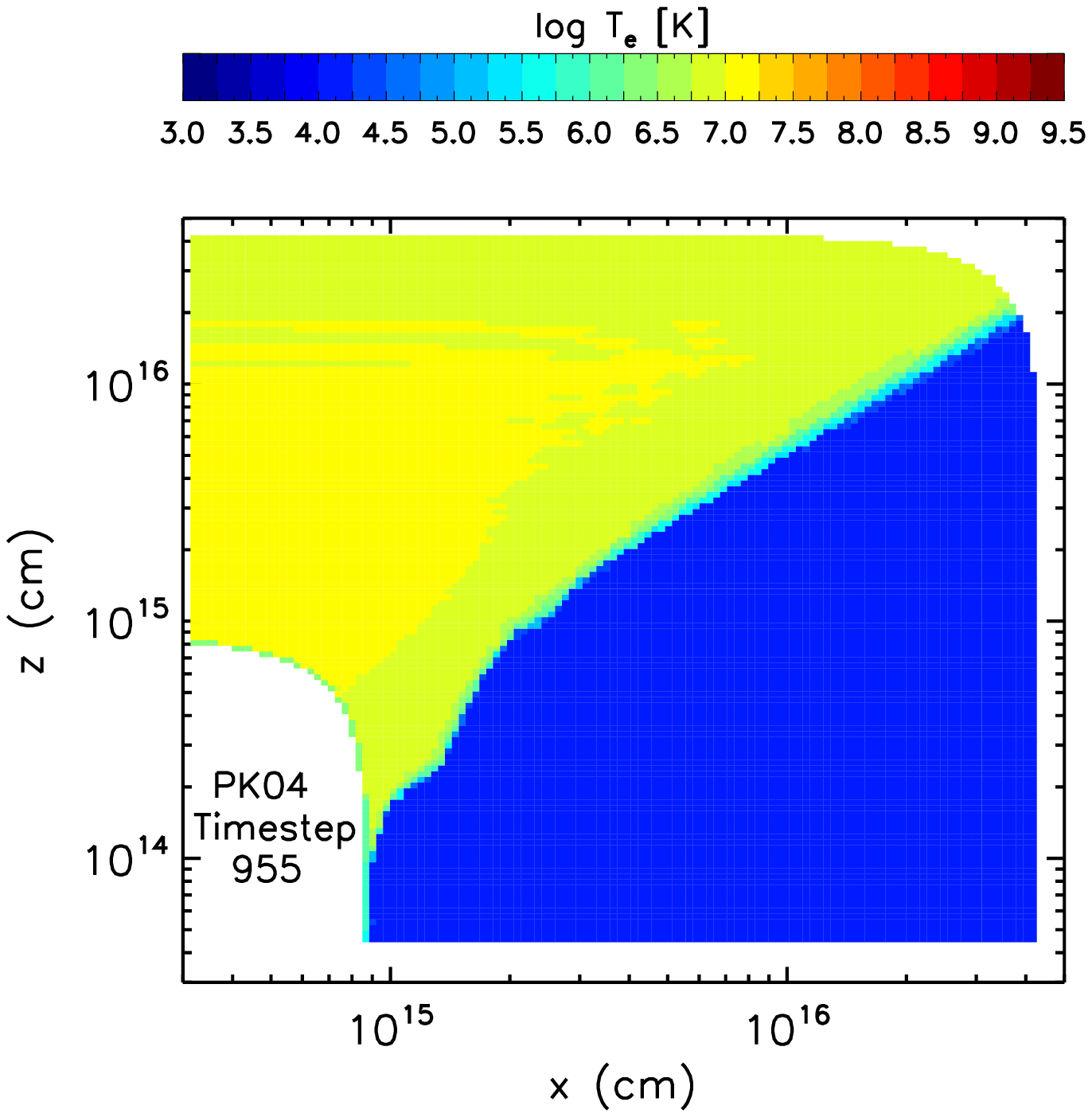, width=7cm,angle=90}\\
\caption{
Distribution of kinetic temperature ($T_{e}$) computed from our Monte
Carlo radiative transfer simulations (top/middle left) and mass
density (top/middle right) for the two snapshots we consider from the
PK04 simulation (timesteps 800 and 955; top and middle rows, respectively.
For 
comparison, we also show the temperature 
obtained for timestep 955 
in the PK04 radiation-hydrodynamics simulations (lower left). 
The lower right panel shows the
temperature distribution obtained with our Monte Carlo code (for
the timestep 955 snapshot) when
heating by scattered/reprocessed radiation is neglected.
Note the logarithmic
axes. The figures show the complete computational domain of the
hydrodynamical simulation (60$r_{g} <  r < 3000 r_{g}$).
A lower temperature limit of $T_{e} = 10^{4}$~K was imposed in the Monte
Carlo radiative transfer simulations.}
\label{fig:temp}
\end{figure*}

The relevant physical and numerical parameters adopted in the
simulations are given in Table~\ref{tab:pars}. Most of these are
carried over or derived from the parameters adopted in the radiation
hydrodynamics simulation (\citealt{proga00}, PK04). For this study we
have selected two snapshots of the wind structure; specifically we
chose the wind conditions from timesteps 800 and
955 from the PK04 simulation. These timesteps were chosen since they
are well-separated in simulation time (timestep 955 is later than
timestep 800 by $\Delta t \sim 5$~years). This allows us to study
two independent realizations of the flow pattern and investigate how
the spectral features should vary on timescales
comparable to the typical interval between repeat observations of
well-known AGN.
Both snapshots show the characteristic structure of the PK04
simulation mentioned in Section~\ref{sect_intro} (a hot polar flow,
warm rapid outflow, cold equatorial flow and a ``failed wind''
region). The relative importance of these structures is slightly
different between the two-snapshots, however: 
this can be seen in the upper
right panels of Figure~\ref{fig:temp} (mass density of
the two-snapshots) and in Figure~\ref{fig:nh}, which compares the
hydrogen column densities for the two snapshots for different
inclination angles, $\theta$ (measured relative to the rotation
(z-)axis.)

For our radiative
transfer simulations we need to fully specify
the X-ray source radiation. We assumed that the
radial extent of the primary emitting region is comparable to the size
of the innermost stable orbit around a Schwarzschild black hole
($r_{em} = 6r_{g}$) and we also assumed that a geometrically
thin/optically thick accretion disk lies in the mid-plane of the
simulation, extending from an inner radius of $r_{d} = 6r_{g}$ to the
outer boundary of the simulation grid. The X-ray source is assumed to
radiate isotropically.

We assume that the primary X-ray radiation source is a
pure power-law across the photon energy interval we simulate ($0.1 <
h\nu < 511$~keV; see Paper~II). We have adopted a photon index of
$\Gamma = 2.1$. This is reasonably typical for
the class of objects for which
disk winds have been discussed as a possible site for the formation of
X-ray spectral features (e.g. $\Gamma \sim 1.8$ for PG1211+143
[\citealt{pounds03}], $\Gamma \sim 2.4$ for Markarian~766 [\citealt{miller07}]
or $\Gamma \sim 2.2$ for MCG-6-30-15 [\citealt{miller08}]).
To estimate the appropriate normalisation of
the primary power-law X-ray emission, we used the source X-ray
luminosity of 0.05~$L_{\mbox{\scriptsize Edd}}$ adopted by
PK04. The role of the X-ray luminosity in the radiation
hydrodynamics simulation is to estimate the ionization parameter
required to compute the
line driving force
multiplier. 
PK04 used the relationship between force multiplier
and ionization parameter derived by \cite{stevens90} which was based
on calculations assuming a bremsstrahlung source
spectrum with temperature $T_{X} = 10$~keV. Therefore, we fixed the normalisation of our input X-ray
luminosity spectrum ($L(E) \propto E^{-\Gamma + 1}$)
  to give the same total X-ray flux (integrated from $1$ to
$20$~keV) as for a bremsstrahlung spectrum ($L(E) = 0.05
  L_{\mbox{\scriptsize Edd}}
\exp(-E/T_{X}) / kT_{X}$). This yields a 2 -- 10~keV luminosity of
$L_{X} = 0.02 L_{\mbox{\scriptsize Edd}}$ which is comparable to
the observationally motivated estimates for $L_{X} /
L_{\mbox{\scriptsize Edd}}$ adopted for the models described in Paper~II.

The Monte Carlo code requires that the total photon
luminosity ($L_{\mbox{\scriptsize bol}}$) 
be specified in order to estimate the role
of Compton cooling in the wind (see
Paper~II). We have adopted
$L_{\mbox{\scriptsize bol}} = 6.5 \times 10^{45}$~ergs~s$^{-1}$ ($ = 0.5 L_{\mbox{\scriptsize Edd}}$)  which
is consistent with the accretion rate ($0.5 \dot{M}_{\mbox{\scriptsize
  Edd}}$) and radiative efficiency ($\eta = 0.06$) adopted by PK04.

\section{Results}
\label{sect:results}

As described in Paper~II, the ionization state and kinetic temperature in the wind
are computed from the radiation field properties via an iterative
sequence of Monte Carlo simulations. We will first describe the
results of this part of the calculation and then present the synthetic spectra.

\subsection{Temperature structure}

Figure~\ref{fig:temp} shows the temperature distribution computed from
the Monte Carlo radiative transfer simulations for the two snapshots
(top and second row, left panels). In both cases the
temperature structure is complex but can be easily understood
as a consequence of the density distributions (top and second row, right panels in
Figure~\ref{fig:temp}) and the central location of the primary X-ray
source.
The low-density polar wind cone is strongly irradiated by the X-ray source
and is therefore both hot and very highly ionized (see
Figure~\ref{fig:ions}, discussed below). 
Below this region lies a warm stream ($T_{e}
\sim 10^6$~K) where the
relatively dense portions of the wind are exposed to the primary
X-rays. Deeper inside, the wind becomes significantly cooler, down to $T_{e}
\simlt$~2~$\times 10^5$~K in the outer regions. The densest
material (comprising the ``failed wind'' region and the atmosphere of
the outer parts of the disk) is not significantly heated by the X-ray
source and remain at the lower temperature boundary imposed in the
simulations ($T_{e} = 10^4$~K).

The computed temperature (and ionization) structure of the disk wind
are noticeably different from those obtained in the original
hydrodynamical simulations of PK04. For comparison, the
temperature distribution for the timestep 955 snapshot obtained from those simulations is
also shown in Figure~\ref{fig:temp} (bottom left panel). There are
several qualitative differences: most importantly, the dense regions
of the outflowing gas are significantly cooler that obtained
here. This can be attributed to the differences in the radiative
transfer schemes employed. In particular, PK04 did not
include scattering -- they assumed that the wind is a pure absorbing
structure. Our Monte Carlo method is considerably more sophisticated
and includes both scattering and true absorption. To illustrate the
importance of this, we also computed the temperature
distribution if only the heating due to direct irradiation is
considered for the same snapshot (i.e. heating by
radiation which is either scattered or reprocessed in the wind is
neglected; see bottom right panel in Figure~\ref{fig:temp}). This
reproduces the low temperatures found in the entire
region shielded by the ``failed wind'' in the radiation hydrodynamics
simulation. Thus our Monte Carlo simulations show that
scattered/reprocessed X-ray radiation can be rather important in
determining the physical conditions in the disk wind, particularly for
material that is shielded from direct radiation by the X-ray source.

\begin{figure}
\epsfig{file=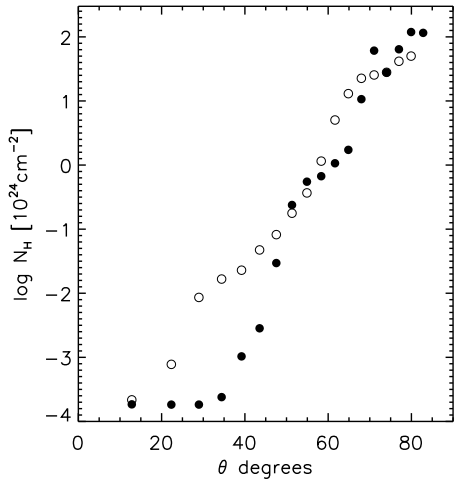, width=7cm,angle=90}
\caption{Hydrogen column density, $N_{H}$ versus inclination angle
  $\theta$ (measured relative to the rotation (z)-axis) for the two
  snapshots from the PK04 simulation for which we compute
  spectra (filled circles for timestep 800, open circles for timestep 955).}
\label{fig:nh}
\end{figure}

There are several other, relatively inconsequential, differences between the
temperature distribution obtained from the Monte Carlo radiative
transfer simulations and the radiation hydrodynamics. 
The polar wind is generally found to be hotter
with the radiation
hydrodynamics code than the Monte Carlo code. This is partly a consequence of
the different forms of the adopted X-ray spectrum. In both codes, much
of this
material is predicted to be close to the Compton temperature $T_{c}$ but PK04
adopted an X-ray temperature of 10~keV (yielding $T_{c} \sim 3 \times
10^7$~K) while the input spectrum of the Monte Carlo simulation
(power-law X-ray spectrum combined with total luminosity $L_{\mbox{\scriptsize bol}}$) gives a lower value of $T_{c} \sim 10^7$~K.
Also, in the radiation hydrodynamics
calculation, the region above the dense
disk wind contains structured regions in which the temperature is much
higher due to shock
heating (interactions between expanding and falling parts of the
flow). Since non-radiative heating is not included in our Monte Carlo
code, such high temperatures are not found in our radiative transfer 
simulations.
We note, however, that
the exactly physical conditions in these polar regions are rather
unimportant for the spectrum since this material is generally optically
thin. Thus the consequences of any discrepancy between the
radiation-hydrodynamics and Monte Carlo radiative transfer simulations
in the low-density polar regions are relatively minor.

\begin{figure*}
\epsfig{file=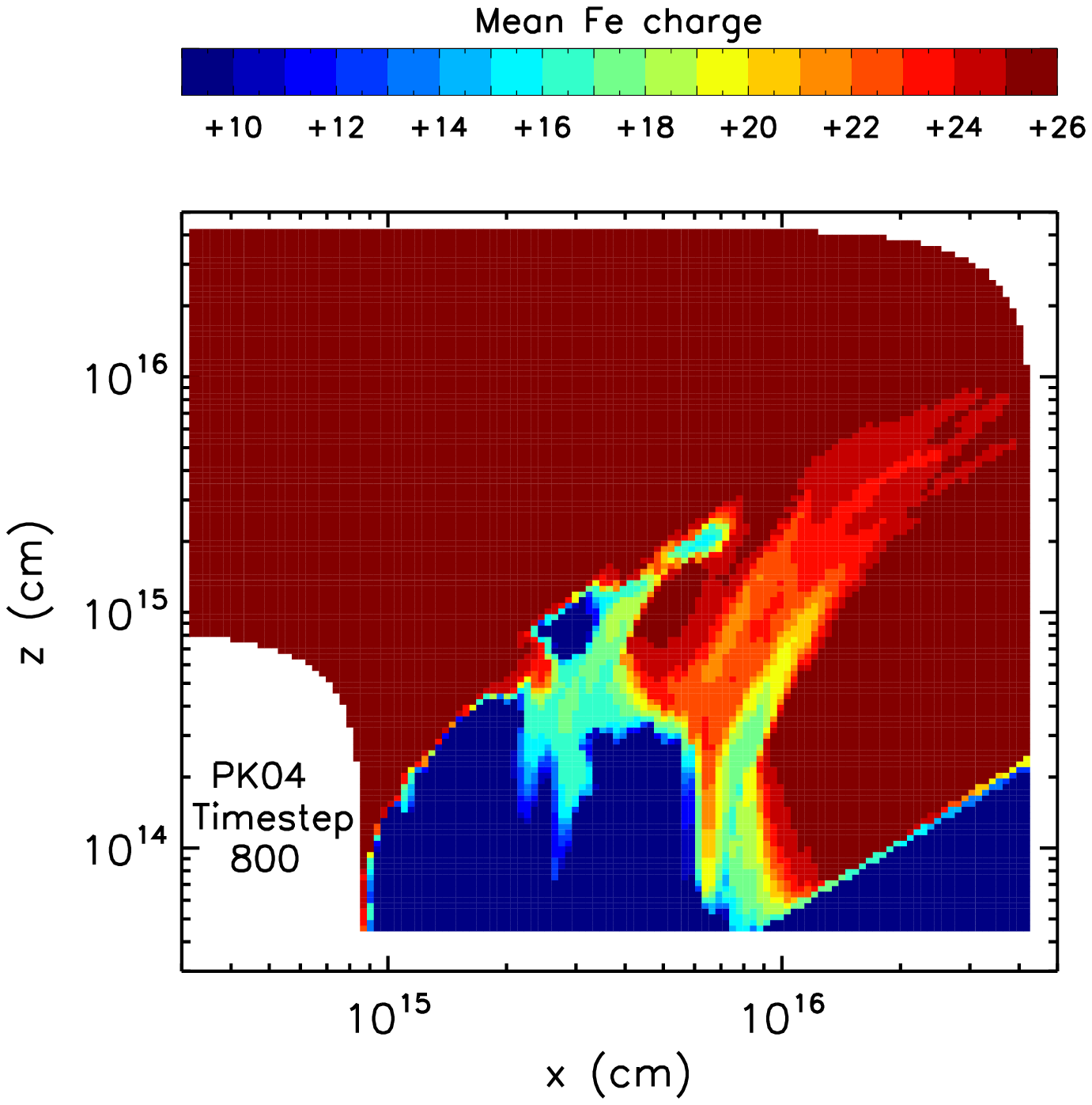, width=7cm,angle=90}
\epsfig{file=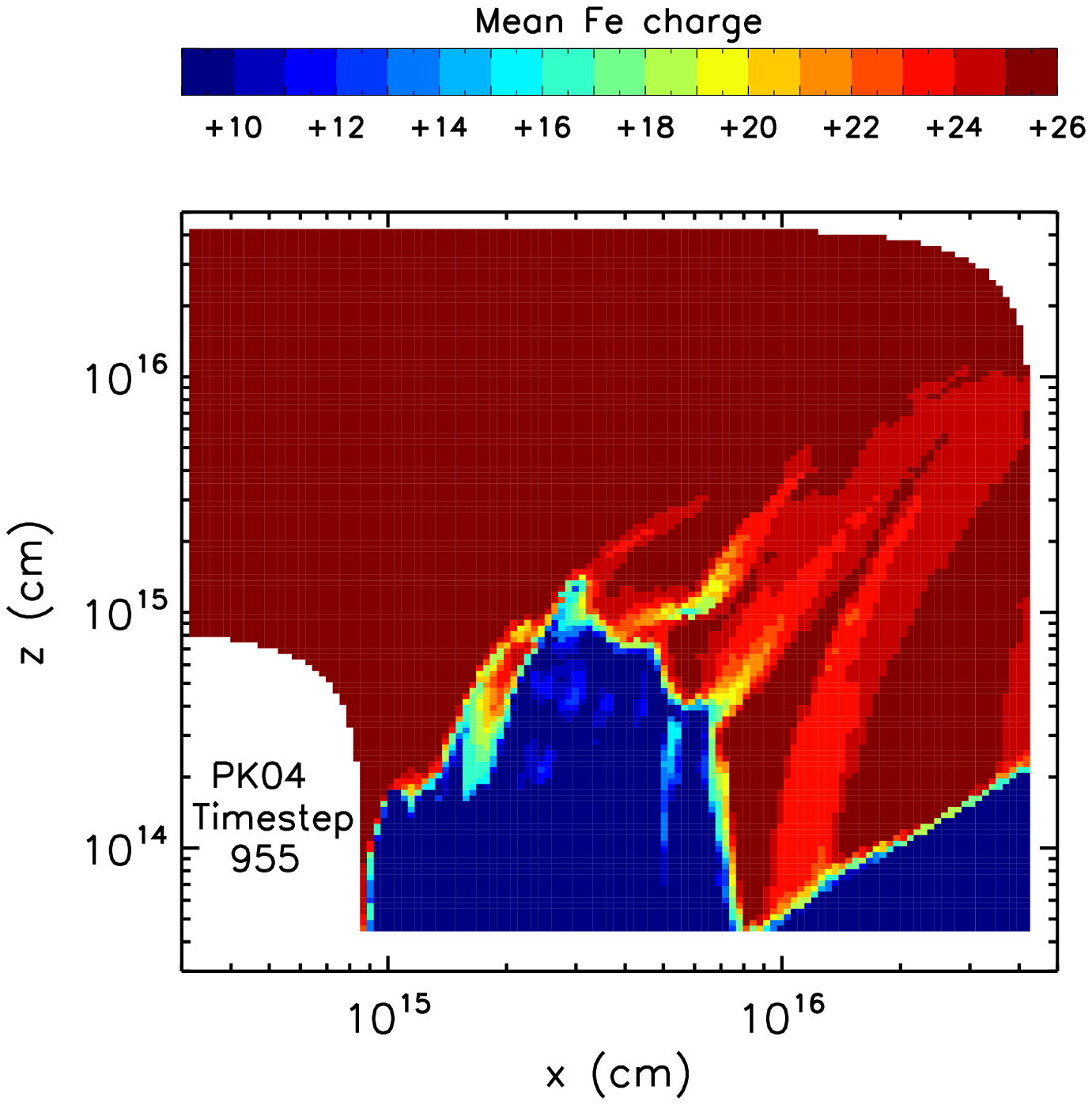, width=7cm,angle=90}
\caption{
The
distribution of mean Fe ionization state for the two snapshots
obtained from the
Monte Carlo code (timesteps 800 and 955 from PK04; left and right, respectively). 
Since we do not include ionization stages of iron below Fe~{\sc x}, the
computed ionization state in the dark blue areas is only an upper
limit on the degree of ionization.}
\label{fig:ions}
\end{figure*}

\subsection{Ionization conditions and Fe~K$\alpha$ forming regions}
\label{sect:ions}

Since the X-rays are responsible for both ionization and heating in
our Monte Carlo simulations, the distribution of ionization state in
the wind is qualitatively
similar to the temperature distribution. This is illustrated in
Figure~\ref{fig:ions} which shows the distribution of
mean iron ionization state computed for both snapshots. In the low
density polar regions the
material is almost fully ionized whilst the denser streams in the flow
are less ionized. Only the dense outer regions of the disk atmosphere
(large radial coordinate) and the ``failed wind'' region are not
significantly ionized.

\begin{figure}
\begin{center}
\epsfig{file=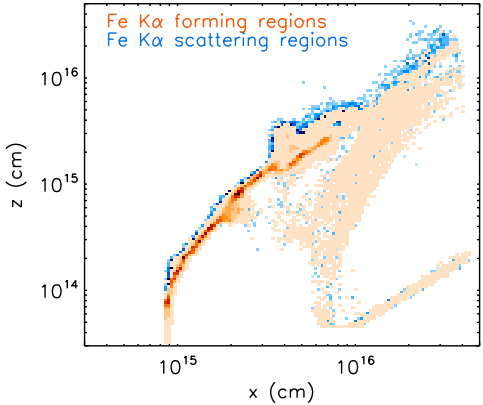, width=7cm,angle=90}
\caption{
Regions responsible for the creation and
subsequent scattering of
the Fe~K$\alpha$ photons which contribute to the spectra for the
snapshot from timestep 800 of the PK04 simulation.
Each grid cell is shaded orange/red if Monte Carlo quanta representing
Fe~K$\alpha$ photons were created in that cell and subsequently
escaped without undergoing any further physical
interactions (except Compton scattering). The intensity of the shading
is proportional to the Fe~K$\alpha$ luminosity of the cell
(ergs~s$^{-1}$; i.e. the volumes of the cells are not taken into account). 
Grid cells are shaded
blue if no Monte Carlo quanta representing Fe~K$\alpha$ photons were
created in that cell but Compton scattering of escaping Fe~K$\alpha$ photons
did occur there. White regions indicate that either no Fe~K$\alpha$
photons were created/scattered there or that any such photons were
subsequently destroyed and so did not contribute to the spectra.
}
\label{fig:cont}
\end{center}
\end{figure}

A noteworthy convenience of the Monte Carlo method is that, by
analyzing the flight paths of quanta, it is relatively easy to
identify which parts of the outflow are
responsible for the formation of particular spectral features. Since
much of the discussion below will be devoted to the Fe~K region, we
have examined the Monte Carlo quanta which contribute to this part of the
spectrum in detail. Specifically, since the Monte Carlo simulation
follows the complete flight paths of the quanta, we can flag each
quantum to identify which
physical processes (e.g. bound-bound transition, Compton scattering etc.)
were responsible for determining its properties (e.g. photon energy,
direction of emission etc.) as it left the computational domain. To
understand the formation of the Fe~K band spectra, we identified all
escaping quanta whose last physical interaction (excluding Compton
scattering) resulted in line emission via an Fe~K$\alpha$
transition. These packets are responsible for
Fe~K$\alpha$ emission features in the spectrum. We then binned these
quanta based on the position of the Fe~K$\alpha$ emission which gave
rise to them. The resulting distribution, which is shown for the PK04
timestep~800 snapshot in
Figure~\ref{fig:cont}, indicates which portions of the flow are
responsible for the creation of the Fe~K$\alpha$ emission
features. Since the emission line profile can be affected by Compton
scattering, we also recorded the positions of the final Compton
scattering event for the quanta flagged as contributing to the
Fe~K$\alpha$ emission. These allow us to identify the parts of the
model in which scattering of the Fe~K$\alpha$ emission takes place --
this region is also indicated in Figure~\ref{fig:cont}. Note that the
plot does not retain any information related to observer
orientation -- all escaping Monte Carlo quanta are included,
independent of their direction of propagation.

Much of the Fe~K$\alpha$ line formation is concentrated in the inner
parts of the flow structure:
roughly 50 per cent of the escaping
Fe~K$\alpha$ photons are created within a radial extent of $r \simlt 2
\times 10^{15}$~cm while $\sim 90$ per cent are formed inside $r \simlt
7 \times 10^{15}$~cm.
Comparing Figure~\ref{fig:cont} to Figures~\ref{fig:temp} and
\ref{fig:ions}, it is apparent that the most intense region of
Fe~K$\alpha$ formation is around the inner surface of the ``failed
wind''. This is expected since
it is here that the X-ray radiation from the central source most
directly strikes relatively dense material from which it can be
reflected. K$\alpha$ formation continues throughout most of the
moderate density outflowing region outside the ``failed wind'',
however, and there is also a reflection component formed from the
dense material near the disk atmosphere (equatorial regions) at large
radii. Shielding by the ``failed wind'' means that this region is not
directly irradiated by the central source but reflection still occurs
thanks to illumination by X-rays scattered in the upper portions of
the wind. 

In the outer wind, there are regions where the ionization
state is too high for significant numbers of Fe~K$\alpha$ photons to
be produced but where Compton scattering of K$\alpha$ photons created
deeper in the wind can still occur. Regions such as this are potentially
critical since Compton scattering in a fast outflow may have a role in
shaping emission line profiles (see e.g. \citealt{laurent07}). 
In our simulations, escaping Fe K$\alpha$ photon packets 
undergo an average of about two 
Compton scatterings between their creation and escape ($\sim 40$
per cent escape with no Compton scattering while $\sim 15$ per cent
undergo five or more Compton scattering events).

\subsection{X-ray spectra}

\subsubsection{General characteristics}

\begin{figure*}
\epsfig{file=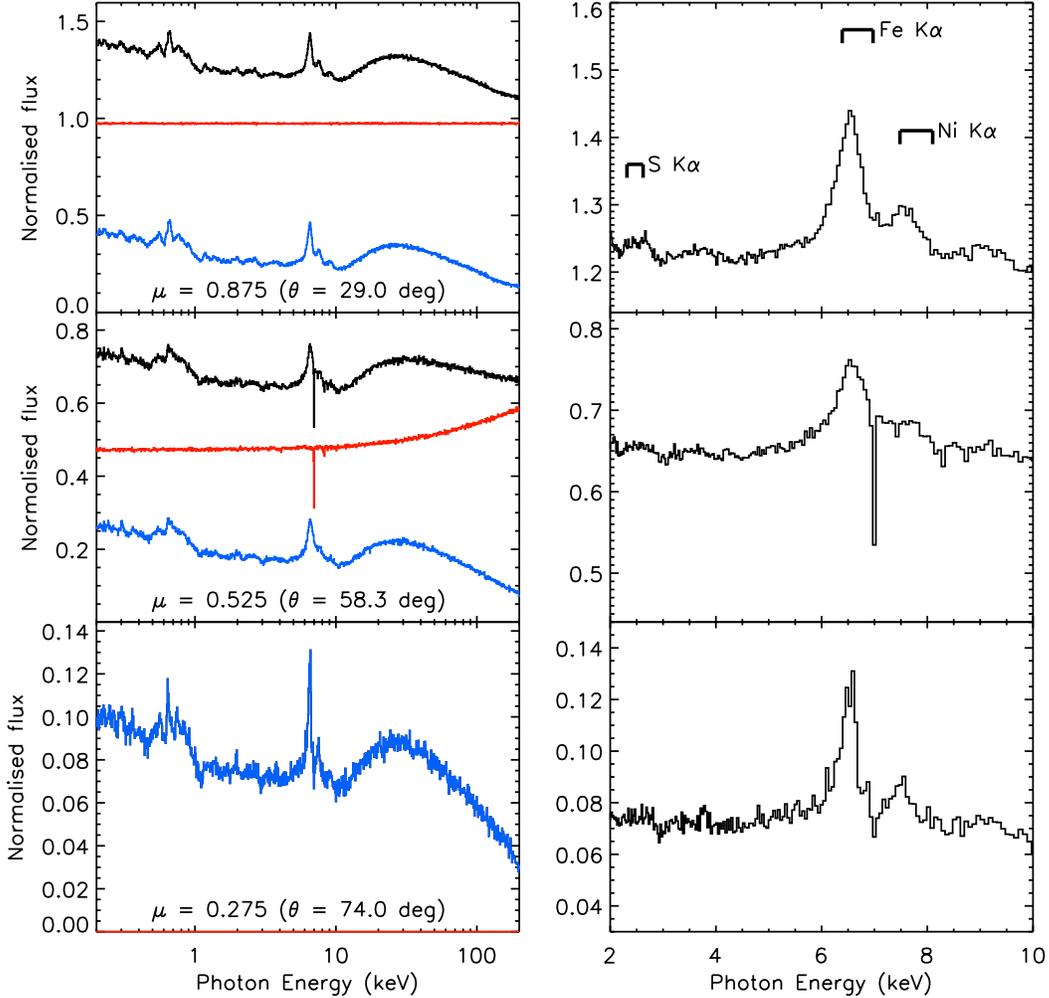, width=14cm,angle=0}
\caption{
Spectra computed for three observer orientations (specified by $\mu =
\cos^{-1} \theta$ where $\theta$ is the inclination of the observer
measured from the z-axis) for timestep 800 of the PK04 simulation.
The left panels show the photon energy range 0.2 -- 200~keV while the
right panels show the 2 -- 10~keV region (including the Fe~K$\alpha$
line) in detail. In the left panels, the total spectrum (black) is
subdivided into contributions from photons which have come directly
from the primary X-ray source (red) and photons which arise from 
scattering/reprocessing in the outflow (blue). All spectra
are normalized to the input primary X-ray spectrum (which was a pure
power-law with photon index $\Gamma = 2.1$). The range of
rest-energies for Fe, Ni and S K$\alpha$ transitions (6.4 -- 6.97~keV,
7.5 -- 8.1~keV and 2.3 -- 2.6~keV, respectively) are indicated in the
upper right panel.}
\label{fig:specs}
\end{figure*}

Given the complex structure of the outflow (see Figure~\ref{fig:temp})
and the high column densities for many lines-of-sight (see
Figure~\ref{fig:nh}) the emergent spectrum is expected to be 
strongly dependent on the observer
inclination angle ($\theta =
\cos^{-1} \mu$). As in Paper~II, we extracted
spectra for twenty different observer orientations, uniformly sampling
the interval $1 > \mu > 0$.
Figure~\ref{fig:specs} shows the spectra computed for
three example observer orientations for the snapshot from timestep 800
of the PK04 simulation. Despite the underlying complexity
of the wind model, these spectra are qualitatively rather similar to
those obtained from the simply-parameterized smooth wind models used
in Paper~II.
The spectra can be broadly classified into three groups, differentiated
by the observer orientation; these are exemplified by the spectra
shown in Figure~\ref{fig:specs} and discussed below.

For relatively pole-on inclinations (first panel of
Figure~\ref{fig:specs}), the direct component of radiation is
largely unattenuated and supplemented by reflected radiation from the
wind. As described in Section~\ref{sect:ions} (see
Figure~\ref{fig:cont}), the most intense reflection occurs at the
inner irradiated surface of the wind (the ``failed wind'' region) and
this gives rise to a reflection component when observed pole
on. Qualitatively similar to reflection contributions in the parametrized disk
wind models used in Paper~II and disk reflection models \citep{ross05},
this reflection component introduces both line emission and a Compton
hump to the spectrum. The emission line features are moderately strong
(the Fe~K$\alpha$ line has equivalent width,
EW $\sim 140$~eV for the upper spectrum in Figure~\ref{fig:cont})
and
significantly broadened by
the Doppler motions of the material (full-width at half-maximum, FWHM
$\sim 700$~eV, determined by Gaussian fitting).
In addition, Compton scattering of
reflection line photons in the outflow causes weak, red-skewed line
wings to develop. This can be
seen in the right-panels of Figure~\ref{fig:specs} which show details
of the spectrum around the strong Fe~K$\alpha$ line.

For moderate inclination angels (second row of
Figure~\ref{fig:specs}), the primary X-ray source is obscured
by the high-ionization, fast outflow component. For these
orientations, Compton scattering is the dominant opacity source along
the line-of-sight although absorption lines associated with high
ionization state material (most clearly, Fe~{\sc xxvi} Ly$\alpha$) are also
imprinted on the transmitted spectrum. 
These features are generally rather narrow but blueshifted from their
rest energies due to the net outflow along the line of sight.
Emission features, typically slightly broader (FWHM $\sim 800$~eV) and
with somewhat more developed red wings than predicted for polar inclinations, 
are also moderately 
strong for these orientations.

At higher inclination, the absorbing column density is larger and the direct
component of the spectrum 
becomes increasingly dominated by the scattered/reprocessed component
(third panel of
Figure~\ref{fig:specs}). This results in a spectrum with
prominent emission lines (Fe~K$\alpha$ EW $\sim 400$~eV) 
and a stronger Compton hump.

In most respects, the spectra we obtain from the two snapshots
lead to qualitatively similar conclusions: in particular, 
the same characteristic spectra features (Compton hump, Fe~K$\alpha$
emission line, blended emission lines at soft energy, blue-shifted
absorption lines) manifest and have similar dependence on the observer
inclination. For fixed orientation,
the emission features also have comparable strength in the two
snapshots, being just
slightly stronger (typically by $\sim 30$ per cent in flux) in the second
snapshot
(recall that the second
snapshot has typically higher column densities which makes
scattered/reprocessed radiation
more dominant).
In detail, however, there are some interesting quantitative 
difference between the absorption line properties of 
the two snapshots, mostly for intermediate observer orientations,
which we will now discuss in greater detail.

\subsubsection{Complexity at intermediate orientations}

\begin{figure*}
\epsfig{file=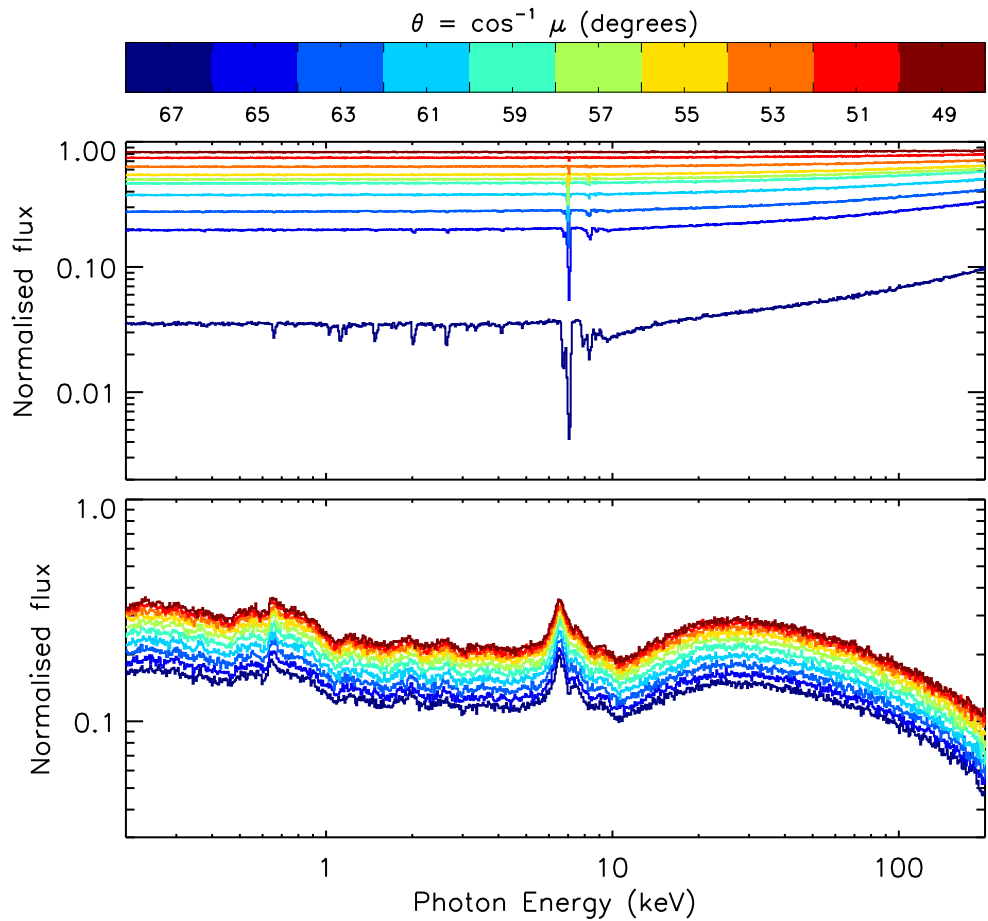, width=13.2cm,angle=90}
\caption{
Direct (upper) and scattered/reprocessed (lower) components of the
spectrum computed for 48~deg $< \theta <$ 67~deg for timestep 800 of
the PK04 simulation (each inclination is
indicated by a different colour; see colour bar in figure). 
The complete model spectrum (see Figure~\ref{fig:specs_zoom}, left) is composed of the sum of these two components.
Note the different ordinate scale used in
the two panels. 
All spectra
are normalized to the input primary X-ray spectrum (which was a pure
power-law with photon index $\Gamma = 2.1$).}
\label{fig:specs_mu}
\end{figure*}

In both snapshots,
the dependence of the spectrum on inclination angle is
relatively modest when the orientation is either high or low. The
sensitivity is strongest for intermediate orientations, mostly because
of the rapid change in the transmission properties of the wind for
inclinations that pass through the complex outflowing structure. 
It is in this range of orientation angles that the most
complex spectra occur 
but also where the least ambiguous outflow
signatures (e.g. narrow, blueshifted absorption lines -- 
see Paper~I) can manifest.
To
map out the spectra in this sensitive regime, we repeated the Monte
Carlo simulations extracting spectra for a finely spaced grid of
twenty inclination angles (with $\Delta \theta = 1$~deg) centred around the
orientation used in the second panel of Figure~\ref{fig:specs}
(48~deg $< \theta <$ 67~deg). A subset of the spectra obtained for this set of
inclination angles in the timestep 800 snapshot is 
shown in Figure~\ref{fig:specs_mu}, divided
into the component of radiation reaching the observer directly from the X-ray
source and that formed by scattering/reprocessing in the disk wind
structure.

\begin{figure*}
\epsfig{file=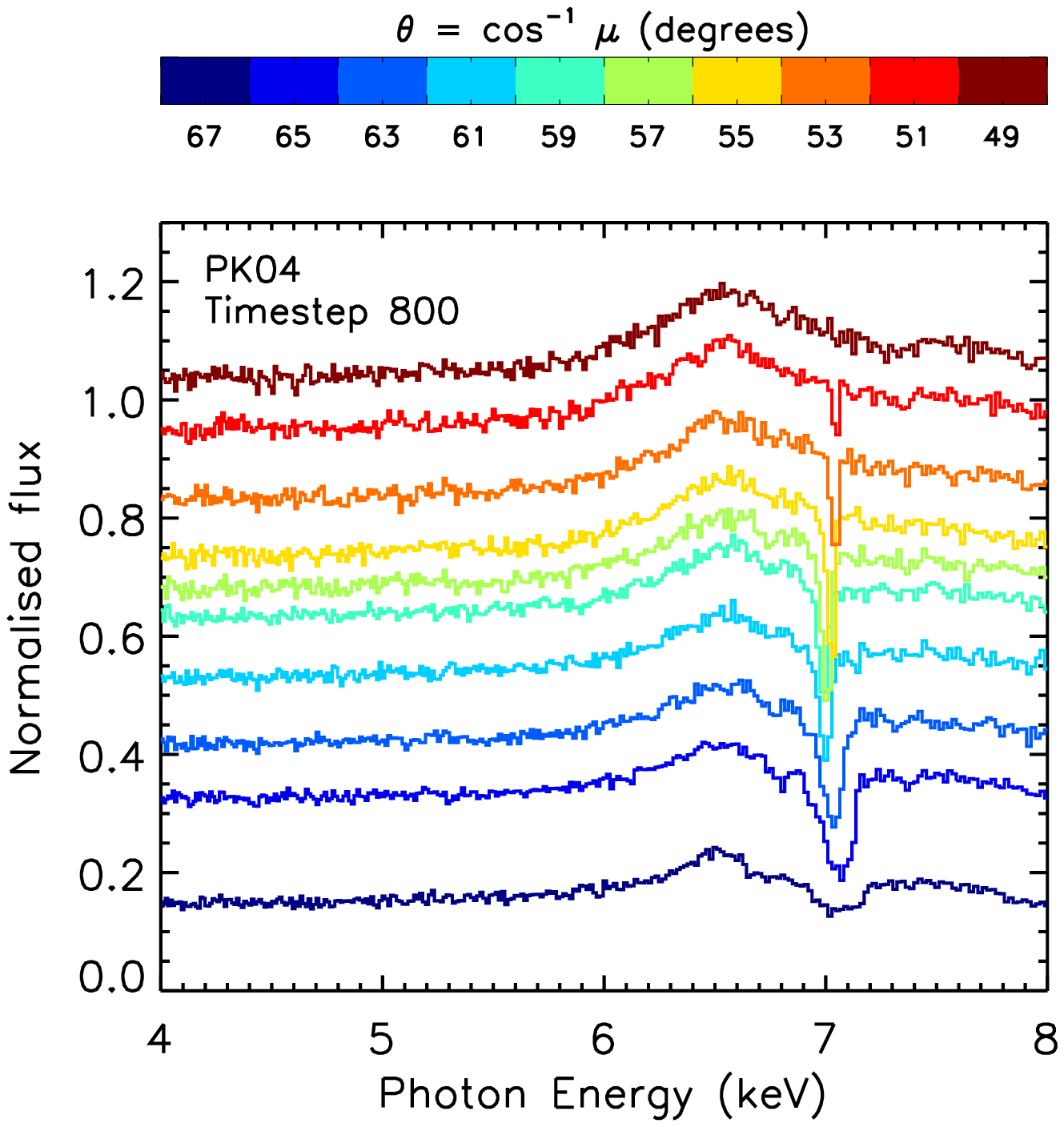, width=8cm,angle=90}
\epsfig{file=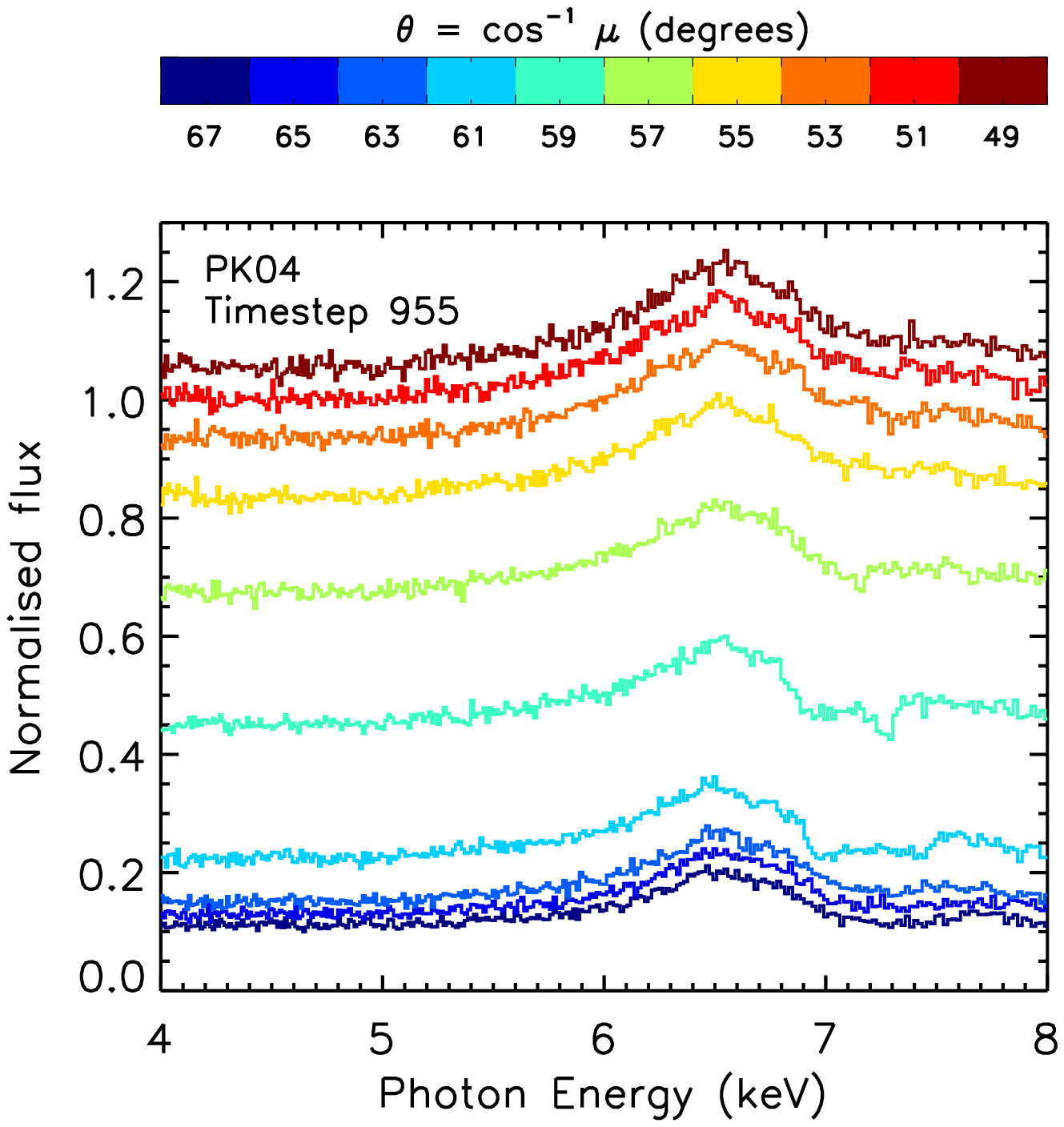, width=8cm,angle=90}
\caption{
Spectra in the Fe~K$\alpha$ region 
for 48~deg $< \theta <$ 67~deg (each inclination is
indicated by a different colour) for the two snapshots (timestep 800
and 955; left and right respectively), which are separated by $\Delta t
\sim 5$~years. 
All spectra
are normalized to the input primary X-ray spectrum (which was a pure
power-law with photon index $\Gamma = 2.1$. The small-scale
fluctuations are due to Monte Carlo noise in the simulations.}
\label{fig:specs_zoom}
\end{figure*}

As expected, the direct component of the spectrum (upper panel in
Figure~\ref{fig:specs_mu}) is a stronger function of inclination
than the scattered/reprocessed component.
This can be readily
understood since the scattered/reprocessed
spectrum is formed by an integration over all positions in the flow
and is therefore relatively insensitive to a small ($\sim 1$~deg)
change in inclination. In contrast, the 
attenuation of the direct component is determined
only by the conditions in the narrow column of gas which obscures the
small (assumed $r_{em} =  6 r_{g}$) X-ray emission region.
Since the properties of the line-of-sight
column change rapidly with inclination, this component of the spectrum
is strongly dependent on $\theta$. Our calculations for the direct component of the
radiation field are in qualitative agreement with the results of
\cite{schurch09}: they demonstrated that strongly
orientation-dependent attenuation is expected in the PK04 simulation
for $50 \mbox{ deg} \simlt \theta \simlt 67 \mbox{ deg}$
(see e.g. their figure 1)\footnote{Note that perfect agreement
  between our calculation of the transmitted spectrum and those of \cite{schurch09} is
  not expected since our calculation includes the effects of
  ionization by scattered light and we have adopted a different
  spectral shape for the source of ionizing radiation. In general, we
  obtain a higher ionization state of the absorbing gas for
  intermediate inclination angles than they found.}.
But since their calculations effectively
include {\it only} the direct component of the radiation field, we
argue that their study overestimates the role of absorption in shaping
the X-ray spectrum: in our emergent
spectra (composed of the the sum of the two components shown in 
Figure~\ref{fig:specs_mu}), scattered/reprocessed light
dominates for many inclinations leading to spectra with less dramatic
absorption features and overall weaker dependence on the observer orientation.

The full spectra (direct plus scattered) around the Fe~K$\alpha$
region for intermediate inclination angles are shown for both of our
snapshots in Figure~\ref{fig:specs_zoom}. This illustrates the 
most striking difference between the spectra obtained from the two
snapshots, namely the behaviour of the blueshifted absorption
lines. Owing to the fairly narrow range of polar angles at which a
fast wind is present in the model, it is only for a modest fraction of
lines-of-sight that clean, sharp absorption
lines appear in our computed spectra for either snapshot. In the
timestep 800 snapshot, an Fe~K$\alpha$ absorption line is present for
a range of inclinations $\Delta \theta \sim 12$~deg. For the later
snapshot, the range is
even smaller, $\Delta \theta \sim 5$~deg.
In our 
previous work (Papers I and II), narrow absorption lines also
only appeared for only a minority of inclinations although they were
generally more common than found here.
This is likely a consequence of the simplicity of the
velocity law adopted in the parametrized models (Papers I and II) -- 
by adopting a
smooth outflow velocity at all points, those models
increase the fraction of
lines-of-sight which pass through moderately opaque material with a
significant component of velocity directed towards the observer. In
contrast, the model considered here has a much more complex velocity
field leading to
a rarer occurrence of clean outflow signatures.

Most importantly, the blueshifted
absorption features are dramatically different between the
two snapshots. In the timestep 800 snapshot, the Fe~K$\alpha$
absorption line not only manifests for a wider range of inclinations
but it is generally significantly sharper, deeper and {\it less} blueshifted compared
to the same inclination in the timestep 955 snapshot. The maximum
Fe~K$\alpha$ blueshift for the first snapshot corresponds to only $\sim 0.015$c
while it is as large as $\sim 0.06$c for the later snapshot. The
largest Fe~K$\alpha$ absorption EW is similar in both snapshots $\sim 70$~eV,
although this occurs for different $\theta$-values ($\theta
\sim 66$~deg for timestep 800 and $\theta \sim 60$~deg for timestep
955). For fixed inclination angle, the absorption EW changes quite
significantly (a factor of two or more) between the two snapshots for
most orientations.
Thus the model predicts that the properties of 
blueshifted absorption features should
not only be a strong function of observer orientation but will also be
significantly variable on timescales comparable to the time difference
between our snapshots ($\Delta t \sim$ 5 years).
We note that the calculated continuum level is also
different between the two snapshots, most obviously
around $\theta \simlt 65$~deg. This is attributable to the larger
column densities for these lines-of-sight in the timestep 955 snapshot
(see Figure~\ref{fig:nh}).

\begin{figure}
\epsfig{file=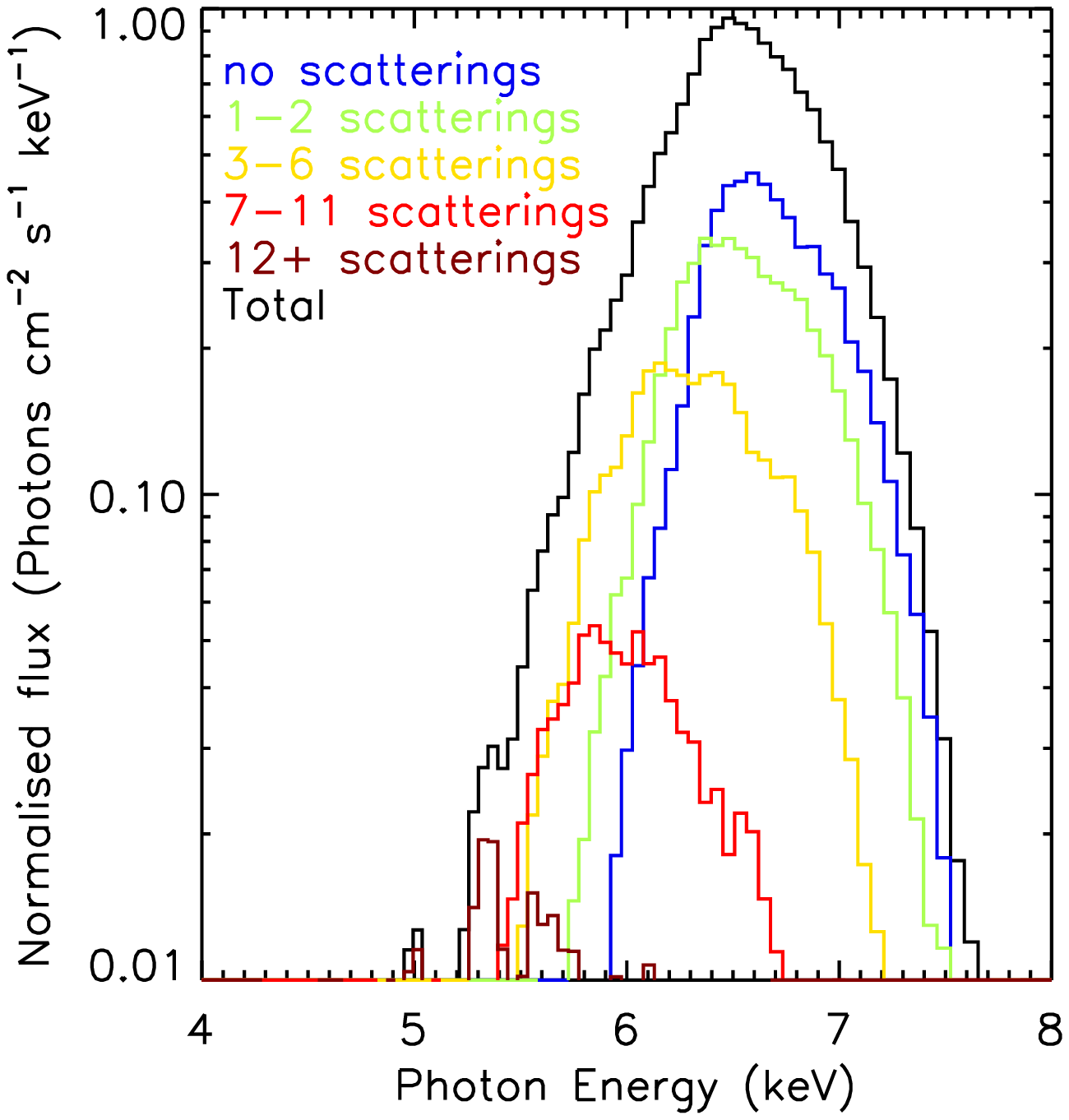, width=8cm,angle=90}
\caption{
The spectrum of escaping Fe~K$\alpha$ photons computed for an observer
inclination around 55~degrees for the timestep 800 snapshot from the
PK04 simulation. The solid black line shows the complete
spectrum of escaping Monte Carlo packets whose last physical
interaction, excluding Compton scattering, was with an Fe~K$\alpha$
transition. The coloured lines show the subdivision of this profile
into contributions from packets which underwent particular numbers of
Compton scattering events after their interaction with an Fe~K$\alpha$
transition. The spectra are normalized to the peak of the line and
have been slightly smoothed to suppress Monte Carlo noise.}
\label{fig:kawing}
\end{figure}

In contrast, the model predicts that emission features (in
particular, Fe~K$\alpha$) should be present in the spectrum for all
observer orientations and that their character will be less
dramatically time-variable (except for $\theta > 75$~deg, 
the Fe~K$\alpha$ emission flux typically
changes by no more than $\sim 30$ per cent between the two snapshots).
The relative insensitivity of the emission line flux 
arises from the fact that they are formed over an extended
region in the flow and are thus less affected by details of the
structure along the observer's line of sight. As noted above, it is at
the same inclination angles for which absorption line features
form that the emission
lines are most intrinsically broad (FWHM $\sim
0.8$~keV) and also where they develop the most noticeably red-skewed wings via
Compton scattering in the flow (see Figure~\ref{fig:kawing}).
We note that the red line wings found in the current simulations are 
somewhat less well-developed than in the simplified models considered in
Papers I and II; this is expected since the fast wind component, in
which Doppler shifts can most effectively give rise to the red-skewed wing,
occupies only a relatively small region of the model. 

\section{Summary, discussion and conclusions}
\label{sect:discussion}

We have applied our Monte Carlo radiative transfer code
(see Papers I and II) to compute
X-ray spectra for snapshots from a
radiation-hydrodynamics simulation of a line-driven AGN disk wind
(PK04). In most important respects, we find that the results
of these simulations support conclusions drawn from our previous studies
of parametrized outflow models. In particular, (1) we find that 
a disk wind should imprint a 
wide range of spectroscopic features in the X-ray band and (2)
the role of the wind in
reflecting/scattering/reprocessing radiation in these simulations 
is {\it at least} as
important as the part it plays in imprinting absorption signatures.

Since we have considered only two snapshots from one model, one should
not expect that the simulated spectra will quantitatively match
observations. Rather the synthetic spectral features should be
interpreted as broadly indicative of those which a disk wind may
cause. This should
guide future studies in which the hydrodynamical modelling 
will be improved and the system parameters explored. 
It should also be borne in mind that the PK04 simulation was
primarily an attempt to build insight to the problem of
line-driven AGN winds and is not expected to capture all the detailed
physics. In particular, since the model is axisymmetric, it will not
correctly describe any small scale clumping in the flow.

In the following sub-sections we first discuss the relevance of our
results to the interpretation of X-ray spectra
(Section~\ref{sect:obs}). We then comment on the comparison of our spectra
to those obtained from the parametrized
models used in Papers~I and II (Section~\ref{sect:param_comp}) before
discussing the implications of our work for the modelling of
line-driven winds (Section~\ref{sect:theory}) and, finally, highlighting some of
the outstanding questions in Section~\ref{sect:more_work}.

\subsection{Implications for the interpretation of X-ray spectra}
\label{sect:obs}

\subsubsection{Absorption lines}

Our first
application of a Monte Carlo code to the modelling of the X-ray
spectra of AGN \citep{sim05b} was motivated by the identification of
narrow, blueshifted absorption lines in the Fe~K band
\citep[e.g.][]{pounds03}. Such features
are the least ambiguous signature of a fast, highly-ionized
outflow. This study confirms that such features can arise in a
plausible disk wind geometry. Our calculations suggest that  narrow,
blueshifted K$\alpha$ absorption lines will manifest in systems with
line-driven disk winds for a limited range of orientations (very
roughly $\Delta
\theta \sim 2 - 10$~deg, centred around moderate inclination angles of
$\theta
\sim 57$~deg). For an isotropic distribution of observed orientations,
this would suggest that only a modest fraction of sources (somewhere
around $\sim 3 - 15$ per cent) should show any such absorption lines. The fraction in the
observed sample of X-ray bright AGN would be expected to be somewhat
higher since very high inclination systems are likely excluded -- but
it should still be a minority. Thus the absence of
this ``smoking gun'' signature of outflow is not evidence against the
presence of a fast wind and the calibration factor required to convert
the fraction of objects with observed narrow line features to the fraction
with powerful disk winds may be large.

Moreover, comparing the spectra obtained from the two
timesteps we considered demonstrates the absorption line properties
should
be significantly variable on timescales comparable to $\Delta t \sim
5$~years. Thus, even if the flow
is persistent, re-observations of particular objects are not expected
to yield very similar absorption line properties over timescales of years: for a fixed observer
orientation, the line blueshifts, widths and equivalent-widths can all
change significantly. This is consistent with observations of
e.g. PG1211+143, an object where blueshifted Fe~K$\alpha$ absorption
has been claimed but in which the absorption properties have varied
between multiple observations made over the course of
the last decade \citep[see
  e.g.][]{pounds03,pounds05,pounds07,pounds09}.
Our current radiative transfer simulations do not constrain the level
of absorption line variability expected on much shorter timescales
(e.g. on the scale of typical X-ray exposures of several
kiloseconds). Detailed consideration of shorter timescales will
require a more extensive, tightly-spaced sequence of snapshots and
investigation of the role played by both small-scale structure in the
flow (which will not be well-represented in the axisymmetric
simulation considered here) and intrinsic, short time-scale
variability of the primary X-ray source. We would like to note,
however, that (i) variations in the wind structure itself occur on a very
wide range of timescales, at least as short as $\sim 10^5$s -- thus we
expect that some degree of variability will also be present on much shorter
timescales than the time difference between the two snapshots used
here; and (ii) short time-scale variability (timescales $\simlt
10^5$~s) is much more likely to manifest in the absorption lines than
any emission features since the light-crossing time for the
simulation ($\sim 0.01$~yr $= 3 \times 10^5$~s) sets a lower limit for
the timescale on which the emission from the outflow should show
significant variability.

\subsubsection{Emission features}

Whilst narrow absorption features appear for only a limited range of
observer orientations, other spectroscopic signatures associated with 
emission/reflection by the wind are predicted to
appear for {\it all} inclination angles:

\begin{enumerate}
\item
A Compton hump is present for all observer
inclinations. This generally causes the spectrum to smoothly rise
across the energy interval $\sim 10 - 25$~keV. At harder energies it
either flattens or, for high inclination angles, turns down. In the
current simulation, the
amplitude of the rise is quite modest, typically several tens per cent
(measured relative to a power-law extrapolated from lower
energies). 

\item
For all observer orientations, the 2 -- 10~keV band shows a fairly broad
Fe K$\alpha$ emission line (and corresponding, but weaker, Ni and S
features). This feature is shaped by fluorescence, recombination and
scattering around the surface of the dense ``failed wind'' region and
in the fast outflow. It can be fairly strong (emission EWs of
at least $\sim$150~eV for most orientations) and can develop a 
modest red-skewed wing due to Compton scattering of line
photons in the outflow. Thus, for a realistic disk wind configuration,
a significant component of strong, broad Fe~K$\alpha$ emission with a
complex line profile can be easily produced.

\item
Below $\sim 1$~keV an additional emission component forms from
multiply blended line and continuum features. These are dominated by
reflection/reprocessing by the abundant elements Fe, O and C and a
modest component of bremsstrahlung emission from the wind. Most of the
emission features are strongly blended although there is always a
fairly prominent O emission feature at 0.7~keV and significant Fe
emission at 0.8~keV. We note that, because it is often a complex
blend,  the exact strength and shape of this
low energy emission will be sensitive to changes in the relative
element abundances (which were fixed to their solar ratios here).

\end{enumerate}

Qualitatively, each of the three classes of feature mentioned above 
is known from
observations of AGN (hard band excesses, Fe K$\alpha$ emission,
excess soft band emission; see e.g. \citealt{turner09}). Given that
there are many simplifying assumptions in the PK04 simulation and that
we have made no attempt to tune the simulation parameters, we find it 
encouraging that
such observable signatures are immediately predicted. 

Although the EWs of our K$\alpha$ emission lines are comparable to
those measured for the strongest narrow K$\alpha$ line
components in AGN, our emission
lines are too broad to account for these: 
\cite{shu10} typically find Fe~K$\alpha$ line cores that have
FWHM $< 10,000$~km~s$^{-1}$ while our lines
  have FWHM $\sim 30,000$~km~s$^{-1}$. The observed narrow line
  cores are therefore more likely to originate in structures further out
  than the main line-forming regions in our simulations. Our synthetic
emission line profiles can be more realistically 
connected with moderately broadened
line components such as reported in MCG-5-23-15 by
\cite{braito07}. For such a case, the agreement in FWHM of the profile
is much closer although the emission EWs obtained from
the current model are too large by a factor of a few. Although
interesting, the amplitude of the rise in flux 
between $\sim 10$ and 25~keV is small compared to
that required to explain {\it Suzaku} observations of hard excesses in
AGN (e.g. PDS~456 \citealt{reeves09}; NCG 4051
\citealt{terashima09}; 1H0419-577 \citealt{turner09a}). To explain
such strong excesses requires absorption by high column-density
optically-thick clumps along the line of sight.
To fully explore this possibility in disk wind models
will require full three-dimensional simulations to be
undertaken in the future (see Section~\ref{sect:more_work}).

We argue that the qualitative properties of the spectra 
obtained from the current
model (with no tuning/exploration of parameter space)
encourages further study of disk wind models. 
We are certainly forced to the strong conclusion that a line-driven disk wind
should imprint several X-ray spectroscopic features in addition to 
``smoking gun'' blueshifted absorption line features. Thus, 
if the fast outflows required to explain blueshifted absorption lines are
associated with a disk wind, it seems
inevitable that the outflow
will also have an important part in shaping many other spectral
features, even when it does not directly
obscure the primary X-ray source region.

\subsection{Comparison to parametrized models}
\label{sect:param_comp}

All the classes of spectral feature described above appeared in our
calculations with simply-parametrized outflow models (see figs. 3 and
4 of Paper~II). 
Qualitatively, the features have comparable strengths and
fairly similar shapes to those obtained with Model~B in Paper~II. The
narrow absorption line features also have approximately the same
maximum blueshifts.
The most apparent quantitative difference is the
relative paucity of narrow absorption features in the spectra computed
here. Also, despite having a wind mass-loss rate ($\dot{M_{w}}$) which is a
significantly smaller fraction of the Eddington accretion rate, the
model presented here predicts emission features which are comparably
strong to those in the parametrized models used in Paper~II
($\dot{M_{w}}
\sim 0.06 \dot{M}_{\mbox{\scriptsize Edd}}$ here but was $\simgt 0.6
\dot{M}_{\mbox{\scriptsize Edd}}$ in the models adopted in
Paper~II). Thus the effective emission measure of the hydrodynamical
wind must be higher than that of the equivalent parametrized model
with the same mass-loss rate. 
In addition, the red-skewed emission line wings 
are typically less
well-developed than in the parametrized models.

These differences can all be attributed 
to the greater complexity of
the hydrodynamical model: in contrast to the parametrized models, it
is not a steady-state flow but has density and velocity
variations that are not simple functions of azimuth and
radius. Moreover, neither the ``failed
wind'' region nor the dense material around the disk at large radii
are captured in the
parametrized model.
Both these regions contribute to the
Fe~K$\alpha$ line emission (see Figure~\ref{fig:cont}; i.e. these
regions increase the effective emission measure of the wind). 

We conclude, therefore, 
that the simply-parameterized models provide a good means by which to
study the influence of a disk wind on the spectrum. In particular, such
models provide a reasonable description of how a successfully launched
outflowing component imprints blueshifted lines and associated broad
emission features from highly ionized material. 
There is, however, an important and obvious caveat with parametric
modelling, namely that it is only useful to the degree that it
reflects the
physical properties of the wind.  In our case, the parametric model is a
poor representation of the "failed wind" component, a component which
does have an affect on the overall spectrum.  One must always be
cautious about  quantitative results derived from a parametric wind
description.  

Taken together,
our studies (Papers I, II and this work) suggest that
models in which a smoothly outflowing component is dominant 
will predict relatively common blueshifted absorption features
(and more strongly-developed red-skewed line wings) 
compared to a flow
with complex topology.
In particular, narrow blueshifted K$\alpha$ absorption is most
likely to be clearly seen in flows with moderate covering fractions
along lines-of-sight that have modest continuum optical depths (see
discussion in Paper~I).
Relative to emission features,
these absorption line features are likely to be rarer in scenarios in
which other flow
components (e.g. the ``failed wind'' in the simulation presented here)
have significant covering fractions.
Thus, there are good prospects that an unbiased census of X-ray
absorption line features could, in the future, place
constraints on the typical geometry required for wind models.

\subsection{Implications for the theory of AGN winds}
\label{sect:theory}

In this study, we have adopted the existing radiation-hydrodynamics
simulation of PK04 and performed detailed radiative
transfer simulations as a post-processing step, assuming that the
underlying density/velocity distributions are appropriate. 
Our
detailed X-ray radiative transfer calculations, however, point to a
potentially important shortcoming of the radiation-hydrodynamics
simulation. 

In the 
radiation-hydrodynamics calculation it was assumed that X-ray
radiative transfer in the wind was a purely absorptive processes.
This leads to the presence of an extended wedge around the base of the
wind which is only weakly ionized since it is shielded from direct
irradiation by a centrally located source. It is in this low
ionization material that the radiative line force due to the uv disk
photons is most effective in accelerating the flow. However, we have
explicitly shown that scattered/reprocessed
radiation has a pivotal role in determining not only the emergent
X-ray spectra but also the ionization/thermal conditions in a disk wind.
In particular, with our multi-dimensional simulations we find that
much of the low density region behind the shielding gas is still
rather significantly ionized thanks to scattered light (see Figure~\ref{fig:temp}).
 
Additional ionization can reduce the line driving, potentially
leading to a weaker wind (or even quenching the wind; see \citealt{proga05}).
We
speculate that the likely consequence of accounting for ionization by
scattered X-rays would be that a
larger fraction of the material would be in the ``failed wind''
component whilst less material will be successfully accelerated to
escape velocity. To investigate this quantitatively will require
further developments of our radiation-hydrodynamics code and
additional simulations. We note, however, that for sources
accreting close to $\dot{M}_{\mbox{\scriptsize Edd}}$, the
formation of the ``failed wind'' component seems unavoidable and our
results suggest that even this
alone is an important structure for the formation of
the X-ray spectrum. Thus it seems unlikely that structures formed due
to the pressure exerted by the uv radiation field of the disk do not
play any part in formation of the X-ray spectrum of AGN accreting
close to the Eddington limit.

\subsection{Further work}
\label{sect:more_work}

Important future improvements to the modelling should include
investigating how the outflowing gas
might continue to affect the spectrum at larger radii and lifting the
assumption of axi-symmetry. A fully three-dimensional flow
could be
more complex and would likely include more density
inhomogeneities (see e.g. right panels of fig. 2 in \citealt{kurosawa09}). This
may influence the observables. In particular, optically thick clumps
might have a role in explaining hard-band excesses (see
e.g. \citealt{reeves09,turner09a}). Since both the wind structure and
primary X-ray source luminosity are likely to be time-variable, the
radiation transport should ultimately be make time-dependent to 
allow for the calculation of
time-lags between spectral features forming in different regions and
departures from thermal/ionization equilibrium. 

Here, our study has been limited to line-driven winds (as simulated by
PK04) but other
processes may also contribute to the launching of AGN outflows:
e.g. magnetically driven or thermal winds from either an irradiated disk
\citep[e.g.][]{luketic10} or an AGN torus
\citep[e.g.][]{dorodnitsyn08a,dorodnitsyn08b,dorodnitsyn09}. 
Thus, in the future, we will extend our studies to consider
alternative wind scenarios and investigate whether there are prospects
to discriminate between possible launching mechanisms based on
spectroscopic observations.
In particular, it will be important to compute spectra based
on a magnetically driven disk wind model. It has been
shown that magnetic fields could drive a disk wind
over a relatively wide range of radii and could
explain some of the properties of AGN outflows 
\citep{blandford82,emmering92,contopoulos94,konigl94,dekool95,bottorff97,bottorff00,everett02,proga03,everett05,fukumura10}.

\section*{Acknowledgments}
 
S.A.S acknowledges the University of Nevada, Las Vegas and the Space
Telescope Science Institute for hosting visits during which part of
this work was carried out.
D.P. acknowledges support provided by the {\it Chandra} award
TM0-11010X issued by the {\it Chandra X-ray Observatory Center}, which is
operated by the Smithsonian Astrophysical Observatory for and on
behalf of NASA under contract NAS 8-39073.
S.A.S. thanks Caroline D'Angelo for several useful discussions.

\bibliographystyle{mn2e}
\bibliography{snoc}

\begin{thebibliography}{}

\bibitem[\protect\citeauthoryear{{Arav}, {Li} \& {Begelman}}{{Arav}
  et~al.}{1994}]{arav94}
{Arav} N.,  {Li} Z.,    {Begelman} M.~C.,  1994, ApJ, 432, 62

\bibitem[\protect\citeauthoryear{{Blandford} \& {Payne}}{{Blandford} \&
  {Payne}}{1982}]{blandford82}
{Blandford} R.~D.,  {Payne} D.~G.,  1982, MNRAS, 199, 883

\bibitem[\protect\citeauthoryear{{Blondin}}{{Blondin}}{1994}]{blondin94}
{Blondin} J.~M.,  1994, ApJ, 435, 756

\bibitem[\protect\citeauthoryear{{Bottorff}, {Korista}, {Shlosman} \&
  {Blandford}}{{Bottorff} et~al.}{1997}]{bottorff97}
{Bottorff} M.,  {Korista} K.~T.,  {Shlosman} I.,    {Blandford} R.~D.,  1997,
  ApJ, 479, 200

\bibitem[\protect\citeauthoryear{{Bottorff}, {Korista} \&
  {Shlosman}}{{Bottorff} et~al.}{2000}]{bottorff00}
{Bottorff} M.~C.,  {Korista} K.~T.,    {Shlosman} I.,  2000, ApJ, 537, 134

\bibitem[\protect\citeauthoryear{{Braito}, {Reeves}, {Dewangan}, {George},
  {Griffiths}, {Markowitz}, {Nandra}, {Porquet}, {Ptak}, {Turner}, {Yaqoob} \&
  {Weaver}}{{Braito} et~al.}{2007}]{braito07}
{Braito} V.,  {Reeves} J.~N.,  {Dewangan} G.~C.,  {George} I.,  {Griffiths}
  R.~E.,  {Markowitz} A.,  {Nandra} K.,  {Porquet} D.,  {Ptak} A.,  {Turner}
  T.~J.,  {Yaqoob} T.,    {Weaver} K.,  2007, ApJ, 670, 978

\bibitem[\protect\citeauthoryear{{Castor}, {Abbott} \& {Klein}}{{Castor}
  et~al.}{1975}]{castor75}
{Castor} J.~I.,  {Abbott} D.~C.,    {Klein} R.~I.,  1975, ApJ, 195, 157

\bibitem[\protect\citeauthoryear{{Contopoulos} \& {Lovelace}}{{Contopoulos} \&
  {Lovelace}}{1994}]{contopoulos94}
{Contopoulos} J.,  {Lovelace} R.~V.~E.,  1994, ApJ, 429, 139

\bibitem[\protect\citeauthoryear{{de Kool} \& {Begelman}}{{de Kool} \&
  {Begelman}}{1995}]{dekool95}
{de Kool} M.,  {Begelman} M.~C.,  1995, ApJ, 455, 448

\bibitem[\protect\citeauthoryear{{Dorodnitsyn} \& {Kallman}}{{Dorodnitsyn} \&
  {Kallman}}{2009}]{dorodnitsyn09}
{Dorodnitsyn} A.,  {Kallman} T.,  2009, ApJ, 703, 1797

\bibitem[\protect\citeauthoryear{{Dorodnitsyn}, {Kallman} \&
  {Proga}}{{Dorodnitsyn} et~al.}{2008a}]{dorodnitsyn08a}
{Dorodnitsyn} A.,  {Kallman} T.,    {Proga} D.,  2008a, ApJL, 675, L5

\bibitem[\protect\citeauthoryear{{Dorodnitsyn}, {Kallman} \&
  {Proga}}{{Dorodnitsyn} et~al.}{2008b}]{dorodnitsyn08b}
{Dorodnitsyn} A.,  {Kallman} T.,    {Proga} D.,  2008b, ApJ, 687, 97

\bibitem[\protect\citeauthoryear{{Emmering}, {Blandford} \&
  {Shlosman}}{{Emmering} et~al.}{1992}]{emmering92}
{Emmering} R.~T.,  {Blandford} R.~D.,    {Shlosman} I.,  1992, ApJ, 385, 460

\bibitem[\protect\citeauthoryear{{Everett}, {K{\"o}nigl} \& {Arav}}{{Everett}
  et~al.}{2002}]{everett02}
{Everett} J.,  {K{\"o}nigl} A.,    {Arav} N.,  2002, ApJ, 569, 671

\bibitem[\protect\citeauthoryear{{Everett}}{{Everett}}{2005}]{everett05}
{Everett} J.~E.,  2005, ApJ, 631, 689

\bibitem[\protect\citeauthoryear{{Friend} \& {Abbott}}{{Friend} \&
  {Abbott}}{1986}]{friend86}
{Friend} D.~B.,  {Abbott} D.~C.,  1986, ApJ, 311, 701

\bibitem[\protect\citeauthoryear{{Fukumura}, {Kazanas}, {Contopoulos} \&
  {Behar}}{{Fukumura} et~al.}{2010}]{fukumura10}
{Fukumura} K.,  {Kazanas} D.,  {Contopoulos} I.,    {Behar} E.,  2010, ApJ,
  715, 636

\bibitem[\protect\citeauthoryear{{K\"{o}nigl} \& {Kartje}}{{K\"{o}nigl} \&
  {Kartje}}{1994}]{konigl94}
{K\"{o}nigl} A.,  {Kartje} J.~F.,  1994, ApJ, 434, 446

\bibitem[\protect\citeauthoryear{{Kurosawa} \& {Proga}}{{Kurosawa} \&
  {Proga}}{2009}]{kurosawa09}
{Kurosawa} R.,  {Proga} D.,  2009, ApJ, 693, 1929

\bibitem[\protect\citeauthoryear{{Laming} \& {Titarchuk}}{{Laming} \&
  {Titarchuk}}{2004}]{laming04}
{Laming} J.~M.,  {Titarchuk} L.,  2004, ApJL, 615, L121

\bibitem[\protect\citeauthoryear{{Laurent} \& {Titarchuk}}{{Laurent} \&
  {Titarchuk}}{2007}]{laurent07}
{Laurent} P.,  {Titarchuk} L.,  2007, ApJ, 656, 1056

\bibitem[\protect\citeauthoryear{{Luketic}, {Proga}, {Kallman}, {Raymond} \&
  {Miller}}{{Luketic} et~al.}{2010}]{luketic10}
{Luketic} S.,  {Proga} D.,  {Kallman} T.~R.,  {Raymond} J.~C.,    {Miller}
  J.~M.,  2010, ArXiv e-prints

\bibitem[\protect\citeauthoryear{{Miller}, {Turner} \& {Reeves}}{{Miller}
  et~al.}{2008}]{miller08}
{Miller} L.,  {Turner} T.~J.,    {Reeves} J.~N.,  2008, A\&A, 483, 437

\bibitem[\protect\citeauthoryear{{Miller}, {Turner}, {Reeves}, {George},
  {Kraemer} \& {Wingert}}{{Miller} et~al.}{2007}]{miller07}
{Miller} L.,  {Turner} T.~J.,  {Reeves} J.~N.,  {George} I.~M.,  {Kraemer}
  S.~B.,    {Wingert} B.,  2007, A\&A, 463, 131

\bibitem[\protect\citeauthoryear{{Pauldrach}, {Puls} \&
  {Kudritzki}}{{Pauldrach} et~al.}{1986}]{pauldrach86}
{Pauldrach} A.,  {Puls} J.,    {Kudritzki} R.~P.,  1986, A\&A, 164, 86

\bibitem[\protect\citeauthoryear{{Pounds} \& {Reeves}}{{Pounds} \&
  {Reeves}}{2007}]{pounds07}
{Pounds} K.~A.,  {Reeves} J.~N.,  2007, MNRAS, 374, 823

\bibitem[\protect\citeauthoryear{{Pounds} \& {Reeves}}{{Pounds} \&
  {Reeves}}{2009}]{pounds09}
{Pounds} K.~A.,  {Reeves} J.~N.,  2009, MNRAS, 397, 249

\bibitem[\protect\citeauthoryear{{Pounds}, {Reeves}, {King}, {Page}, {O'Brien}
  \& {Turner}}{{Pounds} et~al.}{2003}]{pounds03}
{Pounds} K.~A.,  {Reeves} J.~N.,  {King} A.~R.,  {Page} K.~L.,  {O'Brien}
  P.~T.,    {Turner} M.~J.~L.,  2003, MNRAS, 345, 705

\bibitem[\protect\citeauthoryear{{Pounds}, {Reeves}, {King}, {Page}, {O'Brien}
  \& {Turner}}{{Pounds} et~al.}{2005}]{pounds05}
{Pounds} K.~A.,  {Reeves} J.~N.,  {King} A.~R.,  {Page} K.~L.,  {O'Brien}
  P.~T.,    {Turner} M.~J.~L.,  2005, MNRAS, 356, 1599

\bibitem[\protect\citeauthoryear{{Proga}}{{Proga}}{2003}]{proga03}
{Proga} D.,  2003, ApJ, 585, 406

\bibitem[\protect\citeauthoryear{{Proga}}{{Proga}}{2005}]{proga05}
{Proga} D.,  2005, ApJL, 630, L9

\bibitem[\protect\citeauthoryear{{Proga} \& {Kallman}}{{Proga} \&
  {Kallman}}{2004}]{proga04}
{Proga} D.,  {Kallman} T.~R.,  2004, ApJ, 616, 688

\bibitem[\protect\citeauthoryear{{Proga}, {Stone} \& {Kallman}}{{Proga}
  et~al.}{2000}]{proga00}
{Proga} D.,  {Stone} J.~M.,    {Kallman} T.~R.,  2000, ApJ, 543, 686

\bibitem[\protect\citeauthoryear{{Reeves}, {O'Brien}, {Braito}, {Behar},
  {Miller}, {Turner}, {Fabian}, {Kaspi}, {Mushotzky} \& {Ward}}{{Reeves}
  et~al.}{2009}]{reeves09}
{Reeves} J.~N.,  {O'Brien} P.~T.,  {Braito} V.,  {Behar} E.,  {Miller} L.,
  {Turner} T.~J.,  {Fabian} A.~C.,  {Kaspi} S.,  {Mushotzky} R.,    {Ward} M.,
  2009, ApJ, 701, 493

\bibitem[\protect\citeauthoryear{{Ross} \& {Fabian}}{{Ross} \&
  {Fabian}}{2005}]{ross05}
{Ross} R.~R.,  {Fabian} A.~C.,  2005, MNRAS, 358, 211

\bibitem[\protect\citeauthoryear{{Schurch} \& {Done}}{{Schurch} \&
  {Done}}{2007}]{schurch07}
{Schurch} N.~J.,  {Done} C.,  2007, MNRAS, 381, 1413

\bibitem[\protect\citeauthoryear{{Schurch}, {Done} \& {Proga}}{{Schurch}
  et~al.}{2009}]{schurch09}
{Schurch} N.~J.,  {Done} C.,    {Proga} D.,  2009, ApJ, 694, 1

\bibitem[\protect\citeauthoryear{{Shu}, {Yaqoob} \& {Wang}}{{Shu}
  et~al.}{2010}]{shu10}
{Shu} X.~W.,  {Yaqoob} T.,    {Wang} J.~X.,  2010, ApJS, 187, 581

\bibitem[\protect\citeauthoryear{{Sim}}{{Sim}}{2005}]{sim05b}
{Sim} S.~A.,  2005, MNRAS, 356, 531

\bibitem[\protect\citeauthoryear{{Sim}, {Long}, {Miller} \& {Turner}}{{Sim}
  et~al.}{2008}]{sim08}
{Sim} S.~A.,  {Long} K.~S.,  {Miller} L.,    {Turner} T.~J.,  2008, MNRAS, 388,
  611

\bibitem[\protect\citeauthoryear{{Sim}, {Miller}, {Long}, {Turner} \&
  {Reeves}}{{Sim} et~al.}{2010}]{sim10}
{Sim} S.~A.,  {Miller} L.,  {Long} K.~S.,  {Turner} T.~J.,    {Reeves} J.~N.,
  2010, MNRAS, 404, 1369

\bibitem[\protect\citeauthoryear{{Stevens} \& {Kallman}}{{Stevens} \&
  {Kallman}}{1990}]{stevens90}
{Stevens} I.~R.,  {Kallman} T.~R.,  1990, ApJ, 365, 321

\bibitem[\protect\citeauthoryear{{Terashima}, {Gallo}, {Inoue}, {Markowitz},
  {Reeves}, {Anabuki}, {Fabian}, {Griffiths}, {Hayashida}, {Itoh}, {Kokubun},
  {Kubota}, {Miniutti}, {Takahashi}, {Yamauchi} \& {Yonetoku}}{{Terashima}
  et~al.}{2009}]{terashima09}
{Terashima} Y.,  {Gallo} L.~C.,  {Inoue} H.,  {Markowitz} A.~G.,  {Reeves}
  J.~N.,  {Anabuki} N.,  {Fabian} A.~C.,  {Griffiths} R.~E.,  {Hayashida} K.,
  {Itoh} T.,  {Kokubun} N.,  {Kubota} A.,  {Miniutti} G.,  {Takahashi} T.,
  {Yamauchi} M.,    {Yonetoku} D.,  2009, PASJ, 61, 299

\bibitem[\protect\citeauthoryear{{Titarchuk}, {Laurent} \&
  {Shaposhnikov}}{{Titarchuk} et~al.}{2009}]{titarchuk09}
{Titarchuk} L.,  {Laurent} P.,    {Shaposhnikov} N.,  2009, ApJ, 700, 1831

\bibitem[\protect\citeauthoryear{{Turner} \& {Miller}}{{Turner} \&
  {Miller}}{2009}]{turner09}
{Turner} T.~J.,  {Miller} L.,  2009, A\&AR, 17, 47

\bibitem[\protect\citeauthoryear{{Turner}, {Miller}, {Kraemer}, {Reeves} \&
  {Pounds}}{{Turner} et~al.}{2009}]{turner09a}
{Turner} T.~J.,  {Miller} L.,  {Kraemer} S.~B.,  {Reeves} J.~N.,    {Pounds}
  K.~A.,  2009, ApJ, 698, 99

\end{thebibliography}

\label{lastpage}

\end{document}